\documentclass[11pt, oneside]{article}

\pdfoutput=1

\RequirePackage[utf8]{inputenc}
\RequirePackage[T1]{fontenc}
\RequirePackage{xcolor,graphicx}
\RequirePackage{hyperref}
\hypersetup{%
  colorlinks=true,
  breaklinks=true,
  linkcolor=red,
  anchorcolor=black,
  citecolor=brown,
  filecolor=blue,
  menucolor=red,
  urlcolor=red,
}
\RequirePackage[authoryear]{natbib}
\RequirePackage{xr}
\RequirePackage{mathtools, amsfonts, amssymb,amsthm}
\mathtoolsset{showonlyrefs}
\RequirePackage{enumerate}
\RequirePackage{comment}
\RequirePackage{ifthen}
\RequirePackage{enumitem}
\RequirePackage{comment}
\RequirePackage{subcaption}
\RequirePackage{siunitx}
\RequirePackage[ruled,vlined]{algorithm2e}

\RequirePackage{color}
\RequirePackage[normalem]{ulem}

\newcounter{rem}
\newtheorem{remark}[rem]{Remark}

\newcounter{lm}
\newtheorem{lemma}[lm]{Lemma}

\newcounter{scenario}[section]

\newenvironment{results}[1][]{\noindent \textbf{#1} \rmfamily}{\medskip}

\providecommand{\keywords}[1]{\textbf{\textit{Keywords---}} #1}

\newcommand{\EE}{\mathbb{E}} 
\newcommand{\RR}{\mathbb{R}} 
\newcommand{\NN}{\mathbb{N}} 
\newcommand{\XX}{\mathcal{X}} 
\newcommand{\dd}{{\rm d}}

\newcommand{\TT}[1]{\mathcal{T}_{#1}} 
\newcommand{\sLp}[1]{\mathcal{L}^{2}\left(#1\right)} 
\newcommand{\HH}{\mathcal{H}} 

\newcommand{\pointt}{\mathbf{t}} 
\newcommand{\points}{\mathbf{s}} 
\newcommand{\inLp}[2]{\left\langle#1, #2\right\rangle} 
\newcommand{\inR}[2]{\left(#1, #2\right)}
\newcommand{\inRM}[2]{\left(#1, #2\right)_{\mathbf{M}}}
\newcommand{\normLp}[1]{\left|\!\left|#1\right|\!\right|} 
\newcommand{\normR}[1]{\left(\!\left(#1\right)\!\right)} %
\newcommand{\normRM}[1]{\left(\!\left(#1\right)\!\right)_{\mathbf{M}}} %
\newcommand{\inH}[2]{\langle\!\langle#1, #2\rangle\!\rangle}

\newcommand{\inHG}[2]{\langle\!\langle#1, #2\rangle\!\rangle_\Gamma}
\newcommand{\normH}[1]{\left|\!\left|\!\left|#1\right|\!\right|\!\right|}
\newcommand{\normHG}[1]{\left|\!\left|\!\left|#1\right|\!\right|\!\right|_\Gamma}

\newcommand{\Xnp}{X_n^{(p)}} 
\newcommand{\hatXnp}[1]{\widehat{X}_n^{(#1)}} 
\newcommand{\Xp}[1]{X^{(#1)}} 
\newcommand{\mup}[1]{\mu^{(#1)}} 
\newcommand{\hatmup}[1]{\widehat{\mu}^{(#1)}} 
\newcommand{\tildemup}[1]{\widetilde{\mu}^{(#1)}} 
\newcommand{\fp}{f^{(p)}} 
\newcommand{\gp}{g^{(p)}}

\newcommand{\pobs}[1]{\mathrm{#1}} 
\newcommand{\CN}{\mathcal{C}_{\!N}} 
\newcommand{\Gmu}{\pobs{G}_{\!\mu}} 
\newcommand{\OH}{\pobs{O}_{\!\mathcal{H}}} 

\newcommand{\pfea}[1]{\mathsf{#1}} 
\newcommand{\CP}{\mathcal{C}_{\!P}} 
\newcommand{\Gfea}{\pfea{G}_{\!\mu}} 
\newcommand{\OG}{\pfea{O}_{\!\RR}} 

\DeclareMathOperator{\Var}{Var}

\DeclareMathOperator{\bigO}{\mathcal{O}}

\newcommand\restr[2]{{ %
  \left.\kern-\nulldelimiterspace  %
  #1  %
  \vphantom{\big|}  %
  \right|_{#2}  %
}}

\title{On the use of the Gram matrix for multivariate functional principal components analysis}
\author{%
Steven Golovkine\thanks{MACSI, Department of Mathematics and Statistics, University of Limerick, Ireland \href{mailto:steven.golovkine@ul.ie}{steven.golovkine@ul.ie}}
\and
Edward Gunning\thanks{Department of Biostatistics and Epidemiology, University of Pennsylvania, USA \href{mailto:edward.gunning@pennmedicine.upenn.edu}{edward.gunning@pennmedicine.upenn.edu}}
\and
Andrew J. Simpkin\thanks{School of Mathematical and Statistical Sciences, University of Galway, Ireland \href{mailto:andrew.simpkin@nuigalway.ie}{andrew.simpkin@nuigalway.ie}}
\and
Norma Bargary\thanks{MACSI, Department of Mathematics and Statistics, University of Limerick, Ireland \href{mailto:norma.bargary@ul.ie}{norma.bargary@ul.ie}}
}
\date{\today}

\begin{document}
\maketitle

\begin{abstract}
Dimension reduction is crucial in functional data analysis (FDA). The key tool to reduce the dimension of the data is functional principal component analysis. Existing approaches for functional principal component analysis usually involve the diagonalization of the covariance operator. With the increasing size and complexity of functional datasets, estimating the covariance operator has become more challenging. Therefore, there is a growing need for efficient methodologies to estimate the eigencomponents. Using the duality of the space of observations and the space of functional features, we propose to use the inner-product between the curves to estimate the eigenelements of multivariate and multidimensional functional datasets. The relationship between the eigenelements of the covariance operator and those of the inner-product matrix is established. We explore the application of these methodologies in several FDA settings and provide general guidance on their usability.
\end{abstract}

\keywords{Dimension Reduction; Functional Data Analysis; Functional Principal Components; Multivariate Functional Data}

\section{Introduction} 
\label{sec:introduction}

Functional data analysis (FDA) is a statistical methodology for analyzing data that can be characterized as functions. These functions could represent measurements taken over time or space, such as temperature readings over a yearly period or spatial patterns of disease occurrence. The goal of FDA is to extract meaningful information from these functions and to model their behavior. See, e.g., \cite{ramsayFunctionalDataAnalysis2005,horvathInferenceFunctionalData2012,wangFunctionalDataAnalysis2016,kokoszkaSpecialIssueFunctional2017} for some references on FDA.

Functional principal component analysis (FPCA) is an extension of principal component analysis (PCA, a commonly used tool for dimension reduction in multivariate data) to functional data. FPCA was introduced by \cite{karhunenUeberLineareMethoden1947} and \cite{loeveFonctionsAleatoiresStationnaires1945} and developed by \cite{dauxoisAsymptoticTheoryPrincipal1982}. Since then, FPCA has become a prevalent tool in FDA due to its ability to convert infinite-dimensional functional data into finite-dimensional vectors of random scores. These scores are a countable sequence of uncorrelated random variables that can be truncated to a finite vector in practical applications. By applying multivariate data analysis tools to these random scores, FPCA can achieve the goal of dimension reduction while assuming mild assumptions about the underlying stochastic process. FPCA is usually used as a preprocessing step to feed, e.g., regression and classification models. Recently, FPCA has been extended to multivariate functional data, which are data that consist of multiple functions that are observed simultaneously. This extension is referred to as multivariate functional principal component analysis (MFPCA). As for FPCA, a key benefit of MFPCA is that it allows one to identify and visualize the main sources of variation in the multivariate functional data. This can be useful in different applications, such as identifying patterns of movements in sport biomechanics \citep{warmenhovenBivariateFunctionalPrincipal2019}, analyzing changes in brain activity in neuroscience \citep{songSparseMultivariateFunctional2022}, or comparing  countries' competitiveness in economics \citep{krzyskoMultidimensionalEconomicIndicators2022}.

In MFPCA, we seek to decompose the covariance structure of the multivariate functional data into a set of orthogonal basis functions, named the principal components, which capture the main sources of variation in the data. There are multiple approaches to estimate the principal components of a multivariate functional dataset. \cite{ramsayFunctionalDataAnalysis2005} combine the multivariate curves into one big curve and then perform a usual FPCA via an eigendecomposition of the covariance structure. This methodology can only be run for data that are defined on the same unidimensional domain, that exhibit similar amounts of variability and are measured in the same units. \cite{jacquesModelbasedClusteringMultivariate2014a} propose to represent each feature of the multivariate function separately using a basis function expansion. This results in a different set of coefficients for each univariate curve. The eigendecomposition is then run on the matrix of stacked coefficients. To consider the normalization issue of \cite{ramsayFunctionalDataAnalysis2005}, \cite{jacquesModelbasedClusteringMultivariate2014a} and \cite{chiouMultivariateFunctionalPrincipal2014} propose to normalize the data by the standard deviation of the curves at each of the sampling points. \cite{happMultivariateFunctionalPrincipal2018a} extend the estimation of multivariate principal components to functional data defined on different dimensional domains. Their estimation procedure is based on carrying out FPCA on each univariate feature, and then using a weighted combination of the resulting principal components to obtain the multivariate eigencomponents. Finally, \cite{berrenderoPrincipalComponentsMultivariate2011} develop a different method to estimate the eigencomponents as they perform a principal components analysis for each sampling time point.

The key motivation of this paper is to investigate the duality between rows and columns of a data matrix to estimate the eigencomponents of a multivariate functional dataset. The duality between rows and columns of a data matrix is a fundamental concept in classical multivariate statistics \citep{escofierTraitementSimultaneVariables1979,saportaSimultaneousAnalysisQualitative1990}. A data matrix typically represents a set of observations of multiple features, each row corresponds to an individual observation and each column corresponds to an individual feature. The duality between rows and columns refers to the fact that many statistical methodologies can be conducted either on the rows or the columns of the data matrix, and the results will be related to each other. For example, the principal components obtained from a PCA run on the rows of the data matrix are the same as the ones obtained from a PCA run on the columns of the matrix. The choice of method to use, based on criteria such as computational time or data storage needs, is thus left to the user. This concept has been widely studied in multivariate statistics (see, e.g., \cite{pagesMultipleFactorAnalysis2014,hardleAppliedMultivariateStatistical2019}). In the context of functional data, this principle has received limited attention despite being mentioned in the seminal paper of FDA \citep{ramsayWhenDataAre1982a}. \cite{ramsayFunctionalDataAnalysis2005} briefly commented on it in a concluding remark of Chapter~8, while \cite{kneipInferenceDensityFamilies2001} and \cite{benkoCommonFunctionalPrincipal2009} utilized it to compute principal components for dense univariate functional data. \cite{chenQuantifyingInfiniteDimensionalData2017} also employ it to gain computational advantage when univariate functional data are sampled on a very dense grid. To the best of our knowledge, however, there is no available literature on its application to multivariate functional data that are observed on different dimensional domains. Our aim is therefore to investigate this duality for multivariate functional data observed on different dimensional domains and provide guidelines on which method to use in different cases.

The remainder of the paper is organized as follows. In Section~\ref{sec:model}, we define multivariate functional data with components that are observed on possibly different domains. In Section~\ref{sec:geometric_point_of_view_mfpca}, we develop the duality between the observations' space and the functional components' space. The relationship between the eigencomponents of the covariance operator of the multivariate functional datasets and the eigencomponents of the inner-product matrix between the observations is derived in Section~\ref{sec:functional_principal_components_analysis}. Extensive simulations are given in Section~\ref{sec:empirical_analysis}. We also provide guidelines on which method to use with respect to different data characteristics. We present an application related to sports science data from the National Basketball Association (NBA) in Section~\ref{sec:application}. The paper concludes with a discussion and an outlook in Section~\ref{sec:discussion}.



\section{Model} 
\label{sec:model}

The structure of the data we consider, referred to as \emph{multivariate functional data}, is similar to that presented in \cite{happMultivariateFunctionalPrincipal2018a}. The data consist of independent trajectories of a vector-valued stochastic process $X = (\Xp{1}, \dots, \Xp{P})^\top$, $P\geq 1$. (Here and in the following, for any matrix $A$, $A^\top$ denotes its transpose.) For each $1 \leq p \leq P$, let $\TT{p}$ be a rectangle in some Euclidean space $\RR^{d_p}$ with $d_p \geq 1$, e.g., $\TT{p} = [0,1]^{d_p}$. Each coordinate, or feature, $X^{(p)} : \TT{p} \rightarrow \RR$ is assumed to belong to  $\sLp{\TT{p}}$, the Hilbert space of square-integrable real-valued functions defined on $\TT{p}$, having the usual inner product that we denote by $\inLp{\cdot}{\cdot}$, and $\normLp{\cdot}$ the associated norm. Thus $X$ is a stochastic process indexed by $\pointt = (t_1, \ldots, t_P)$ belonging to the $P-$fold Cartesian product $\TT{} : =\TT{1} \times \cdots \times \TT{P}$ and taking values in the $P-$fold Cartesian product space $\HH \coloneqq \sLp{\TT{1}} \times \dots \times \sLp{\TT{P}}$. 

We consider the function $\inH{\cdot}{\cdot} : \HH \times \HH \rightarrow \RR$,
\begin{equation}\label{eq:innerprodH}
    \inH{f}{g} \coloneqq \sum_{p=1}^{P} \inLp{\fp}{\gp} = \sum_{p=1}^{P}\int_{\TT{p}} \fp(t_p)\gp(t_p) \dd t_p, \quad f, g \in \HH.
\end{equation}
$\HH$ is a Hilbert space with respect to the inner product $\inH{\cdot}{\cdot}$\citep{happMultivariateFunctionalPrincipal2018a}. We denote by $\normH{\cdot}$, the norm induced by $\inH{\cdot}{\cdot}$. Let $\mu : \TT{} \rightarrow \HH$ denote the mean function of the process $X$, $\mu(\pointt) \coloneqq \EE(X(\pointt)),\,\pointt \in \TT{}$. Let $C$ denote the $P \times P$ matrix-valued covariance function which, for $\points, \pointt \in \TT{}$, is defined as
\begin{equation}\label{eq:covariance_function}
    C(\points, \pointt) \coloneqq \EE\left(\{X(\points) - \mu(\points)\}\{X(\pointt) - \mu(\pointt)\}^{\top}\right), \quad \points, \pointt \in \TT{}.
\end{equation}
More precisely, for $1 \leq p, q \leq P$, the $(p, q)$th entry of the matrix $C(\points, \pointt)$ is the covariance function between the $p$th and the $q$th features of the process $X$:
\begin{equation}\label{eq:covariance_function_components}
    C_{p, q}(s_p, t_q) \coloneqq \EE\left(\{\Xp{p}(s_p) - \mup{p}(s_p)\}\{\Xp{q}(t_q) - \mup{q}(t_q)\}\right), \quad s_p \in \TT{p}, t_q \in \TT{q}.
\end{equation}
Let $\Gamma : \HH \rightarrow \HH$ denote the covariance operator of $X$, defined as an integral operator with kernel $C$. That is, for $f \in \HH$ and $\pointt \in \TT{}$, the $p$th feature of $\Gamma f(\pointt)$ is given by
\begin{equation}\label{eq:covariance_operator_components}
    (\Gamma f)^{(p)}(t_p) \coloneqq \inH{C_{p, \cdot}(t_p, \cdot)}{f(\cdot)} = \inH{C_{\cdot, p}(\cdot, t_p)}{f(\cdot)}, \quad t_p \in \TT{p}.
\end{equation}

\begin{figure}
    \centering
    \includegraphics[scale=0.9]{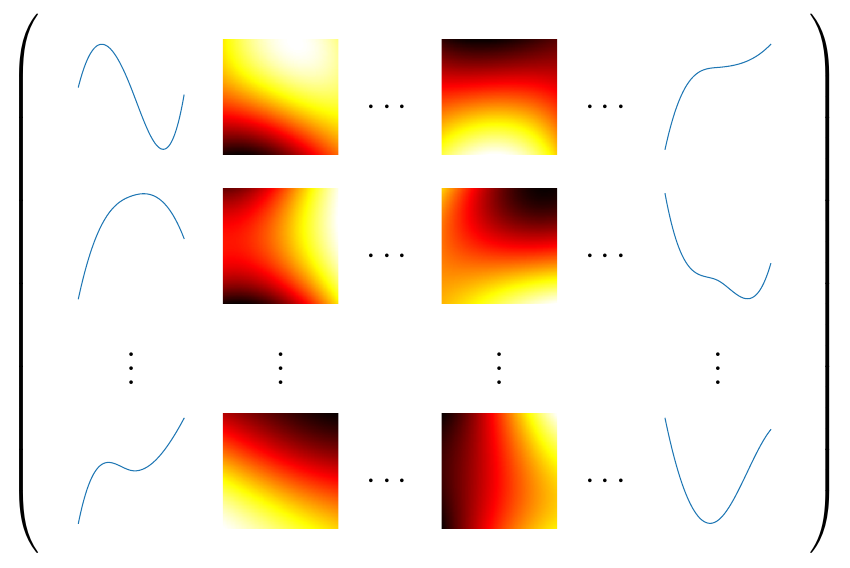}
    \caption{Functional data matrix, adapted from \cite{berrenderoPrincipalComponentsMultivariate2011}.}
    \label{fig:data_matrix}
\end{figure}

Let us consider a set of $N$ independent multivariate curves $\XX = \{X_n\}_{1 \leq n \leq N}$ generated as a random sample of the $P$-dimensional stochastic process $X$ with continuous trajectories. The data can be viewed as a table with $N$ rows and $P$ columns where each entry is a curve, potentially on a multidimensional domain (see Figure~\ref{fig:data_matrix}). Each row of this matrix represents an observation; while each column represents a functional feature. At the intersection of row $n$ and column $p$, we thus have $\Xnp$ which is the curve that concerns the (functional) feature $p$ for the individual $n$. For $n \in \{1, \dots, N\}$, each observation $n$ is attributed the weight $\pi_n$ such that $\sum_n \pi_n = 1$, e.g., $\pi_n = 1/N$. For the set $\mathcal{X}$, the inner-product matrix, also called the Gram matrix, $\mathbf{M}$ is defined as a matrix of size $N \times N$ with entries
\begin{equation}\label{eq:gram_mat}
    \mathbf{M}_{nn^\prime} = \sqrt{\pi_n \pi_{n^{\prime}}}\inH{X_n - \mu}{X_{n^\prime} - \mu}, \quad n, n^\prime = 1, \dots, N.
\end{equation}
This matrix is symmetric, positive definite, and interpretable as a proximity matrix, each entry being the similarity between the weighted observations.

This setting can be generalized to incorporate a general weighting scheme, as \cite{chiouMultivariateFunctionalPrincipal2014} or \cite{happMultivariateFunctionalPrincipal2018a}. We consider the function $\inH{\cdot}{\cdot}_w~:~\HH \times \HH \rightarrow \RR$,
\begin{equation}\label{eq:innerprodH_weight}
    \inH{f}{g}_w \coloneqq \sum_{p=1}^{P} w_p\inLp{\fp}{\gp} = \sum_{p=1}^{P}w_p\int_{\TT{p}} \fp(t_p)\gp(t_p) \dd t_p, \quad f, g \in \HH.
\end{equation}
Let $\Gamma_w : \HH \rightarrow \HH$ denote the weighted covariance operator of $X$, defined as an integral operator with kernel $C$. That is, for $f \in \HH$ and $\pointt \in \TT{}$, the $p$th feature of $\Gamma_w f(\pointt)$ is given by
\begin{equation}\label{eq:covariance_operator_components_weight}
    (\Gamma_w f)^{(p)}(t_p) \coloneqq \inH{C_{p, \cdot}(t_p, \cdot)}{f(\cdot)}_w = \inH{C_{\cdot, p}(\cdot, t_p)}{f(\cdot)}_w, \quad t_p \in \TT{p}.
\end{equation}
Similarly, we define the weigthed inner-product matrix $\mathbf{M}_w$, the matrix of size $N \times N$ with entries
\begin{equation}\label{eq:gram_mat_weights}
    \mathbf{M}_{w, nn^\prime} = \sqrt{\pi_n \pi_{n^{\prime}}}\inH{X_n - \mu}{X_{n^\prime} - \mu}_w, \quad n, n^\prime = 1, \dots, N.
\end{equation}
\begin{remark}
The observation weights $\pi_n$ and the feature weights $w_p$ are not the same and should not be confused. The observation weights are used to give more (or less) importance to specific observations, while the feature weights control for different levels of variation for the features. A discussion on the feature weights is provided in Section~\ref{sub:on_centering_and_reducing}.    
\end{remark}

\subsection{Inference} 
\label{sub:inference}

If we observe the set $\mathcal{X}$, the ideal estimators of the mean and covariance function are
\begin{equation}\label{eq:perfect_estimator}
    \widetilde{\mu}(\pointt) = \sum_{n = 1}^N \pi_n X_n(\pointt), \quad\text{and}\quad \widetilde{C}(\points, \pointt) =  \sum_{n = 1}^N \pi_n \{X_n(\points) - \widetilde{\mu}(\points)\}\{X_n(\pointt) - \widetilde{\mu}(\pointt)\}^\top, \quad \points, \pointt \in \TT{}.
\end{equation}
Similarly, one could estimate the inner-product $\mathbf{M}$ by replacing the mean by their estimators. The resulting matrix $\widetilde{\mathbf{M}}$ has entries
\begin{equation}\label{eq:perfect_gram_estimator}
    \widetilde{\mathbf{M}}_{nn^\prime} = \sqrt{\pi_n \pi_{n^{\prime}}}\sum_{p = 1}^P \int_{\TT{p}}\{\Xnp(t_p) - \tildemup{p}(t_p)\}\{X_{n^\prime}^{(p)}(t_p) - \tildemup{p}(t_p)\} \dd t_p, \quad n, n^\prime = 1, \dots, N.
\end{equation}
In many applications, the elements of the set $\mathcal{X}$ are observed with error and on a finite grid of points in $\TT{}$. For each $1 \leq n \leq N$, and given a vector of positive integers $\mathsf{M}_n = (\mathsf{M}_n^{(1)}, \dots, \mathsf{M}_n^{(P)})$, let $\mathrm{T}_{n, \mathsf{m}} = (\mathrm{T}_{n, m_1}^{(1)}, \dots, \mathrm{T}_{n, m_P}^{(P)}), 1 \leq m_p \leq \mathsf{M}_n^{(p)}, 1 \leq p \leq P$, be the random observation times for the curve $X_n$. These times are obtained as independent realizations of a random variable $\mathbf{T}$ taking values in $\TT{}$. The vectors $\mathsf{M}_1, \dots, \mathsf{M}_N$ represent an independent sample of an integer-valued random vector $\mathsf{M}$ with known expectation. We assume that the realizations of $X$, $\mathsf{M}$ and $\mathbf{T}$ are mutually independent. The observations associated with an observation $X_n$ consist of the pairs $(Y_{n, \mathsf{m}}, \mathrm{T}_{n, \mathsf{m}}) \in \mathbb{R}^P \times \TT{}$, where $\mathsf{m} = (m_1, \dots, m_P), 1 \leq m_p \leq \mathsf{M}_n^{(p)}, 1 \leq p \leq P$ and $Y_{n, \mathsf{m}}$ is defined as
\begin{equation}\label{eq:model_error}
    Y_{n, \mathsf{m}} = X_n(\mathrm{T}_{n, \mathsf{m}}) + \varepsilon_{n, \mathsf{m}}, \quad 1 \leq n \leq N,
\end{equation}
with the $\varepsilon_{n, \mathsf{m}}$ being independent realizations of a centered error random vector $\varepsilon \in \RR^{P}$ with finite variance $\sigma^2\mathbf{1}_P \in \RR^{P \times P}$ where $\sigma^2 = (\sigma^2_1, \dots, \sigma^2_P) \in \RR^{P}$.

Let $\widehat{X}_n$ be a suitable estimator of $X_n$ applied to the pairs $(Y_{n, \mathsf{m}}, \mathrm{T}_{n, \mathsf{m}})$, such as P-splines \cite[e.g.][]{eilersTwentyYearsPsplines2015} or local polynomials \cite[e.g.][]{fanLocalPolynomialModelling1996}. We define the estimator of the mean function as
\begin{equation}
    \widehat{\mu}(\pointt) = \sum_{n = 1}^N \pi_n \widehat{X}_n(\pointt), \quad \pointt \in \TT{}.
\end{equation}
Concerning the estimation of the covariance of the $p$th feature, we distinguish the diagonal from the non-diagonal points. The estimation of the covariance function of the non-diagonal points is defined as
\begin{equation}\label{eq:cov_estimation}
    \widehat{C}_{p, p}(s_p, t_p) = \sum_{n = 1}^N \pi_n\left(\{\hatXnp{p}(s_p) - \hatmup{p}(s_p)\}\{\hatXnp{p}(t_p) - \hatmup{p}(t_p)\}\right), \quad s_p \neq t_p, \quad s_p, t_p \in \TT{p}.
\end{equation}
The variance function $C_{p, p}(s_p, t_p)$ induces a singularity when estimating the covariance function $C_{p, p}(\cdot, \cdot)$ \cite[see][]{yaoFunctionalDataAnalysis2005,zhangSparseDenseFunctional2016}. We define the estimator of the diagonal of the covariance as
\begin{equation}
    \widehat{C}_{p, p}(s_p, s_p) = \sum_{n = 1}^N \pi_n\{\hatXnp{p}(s_p) - \hatmup{p}(s_p)\}^2 - \sigma_p^2, \quad s_p \in \TT{p}.
\end{equation}
This singularity does not appear in the estimation of the cross-covariance function. We define the estimator of the cross-covariance function between the $p$th and $q$th features by
\begin{equation}
    \widehat{C}_{p, q}(s_p, t_q) = \sum_{n = 1}^N \pi_n\left(\{\hatXnp{p}(s_p) - \hatmup{p}(s_p)\}\{\hatXnp{q}(t_q) - \hatmup{q}(t_q)\}\right), \quad s_p \in \TT{p}, t_q \in \TT{q}.
\end{equation}
The estimator of the inner-product matrix can be defined by replacing the curves by their estimators. Following \cite{benkoCommonFunctionalPrincipal2009} and \cite{grithFunctionalPrincipalComponent2018}, we define the estimator of the inner-product matrix as
\begin{equation}
    \widehat{\mathbf{M}}_{nn^\prime} = \sqrt{\pi_n \pi_{n^{\prime}}}\sum_{p = 1}^P \int_{\TT{p}}\{\hatXnp{p}(t_p) - \widehat{\mu}^{(p)}(t_p)\}\{\widehat{X}^{(p)}_{n^\prime}(t_p) - \widehat{\mu}^{(p)}(t_p)\} \dd t_p, \quad n, n^\prime = 1, \dots, N.
\end{equation}
As for the covariance estimation, the model implies a bias on the diagonal term of the inner-product matrix. A correction for the diagonal of the Gram matrix is given by
\begin{equation}
    \widehat{\mathbf{M}}_{nn} = \pi_n\sum_{p = 1}^P \left\{\int_{\TT{p}}\{\hatXnp{p}(t_p) - \widehat{\mu}^{(p)}(t_p)\}^2 \dd t_p - \sigma_p^2 \right\}, \quad n = 1, \dots, N.
\end{equation}
For the estimation of the variances $\sigma_p^2$, we let the reader refers to \cite{hallVarianceEstimationNonparametric1990} and \cite{hallAsymptoticallyOptimalDifferenceBased1990}.



\section{On the geometry of multivariate functional data} 
\label{sec:geometric_point_of_view_mfpca}

\subsection{Duality diagram} 
\label{sub:duality_diagram}

The distinction between the space of rows of a matrix as a sample from a population and the space of columns as the fixed variables on which the observations were measured has been explained in \cite{holmesMultivariateDataAnalysis2008} and \cite{delacruzDualityDiagramData2011} for multivariate data. We propose to define a duality diagram in the context of multivariate functional data. Consider the data matrix defined by the set $\mathcal{X}$. We define an operator $L_X : \HH \rightarrow \RR^N$ by
\begin{equation}
    L_X: f \mapsto \begin{pmatrix}
        \sqrt{\pi_1}\inH{X_1 - \mu}{f} \\
        \vdots \\
        \sqrt{\pi_N}\inH{X_N - \mu}{f}
    \end{pmatrix}.
\end{equation}
Using the linearity of the inner-product $\inH{\cdot}{\cdot}$ and vectors, the operator $L_X$ is linear. Define the operator $L^\star_X : \RR^N \rightarrow \HH$ as
\begin{equation}
    L^\star_X: u \mapsto \begin{pmatrix}
       \sum_{n = 1}^N \sqrt{\pi_n} u_n \{X_n^{(1)}(t_1) - \mup{1}(t_1)\} \\ 
       \vdots \\ 
       \sum_{n = 1}^N \sqrt{\pi_n} u_n \{X_n^{(P)}(t_P) - \mup{P}(t_P)\}
    \end{pmatrix}.
\end{equation}

\begin{figure}
    \centering
    \includegraphics[scale=1.2]{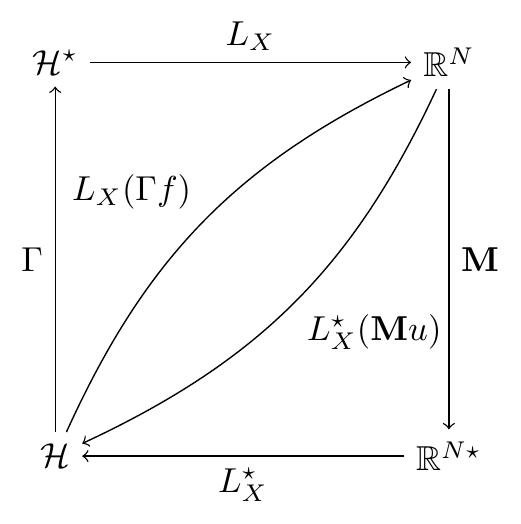}
    \caption{Duality diagram between the spaces $\mathcal{H}$ and $\mathbb{R}^N$. The operator $L_X$ and its adjoint $L^\star_X$ are linear operators. The covariance operator $\Gamma$ and the matrix $\mathbf{M}$ define geometries in $\mathcal{H}$ and $\mathbb{R}^N$ respectively. The space $\HH^\star$ (resp. $\RR^{N \star}$) is the dual space of $\HH$ (resp. $\RR^N$).}
    \label{fig:duality_diagram}
\end{figure}

\begin{lemma}\label{lm:adjoint}
    $L^\star_X$ is the adjoint operator of the linear operator $L_X$.
\end{lemma}
The proof of Lemma \ref{lm:adjoint} is given in Appendix~\ref{sec:derivation_of_the_inertia_of_the_clouds}. As an adjoint operator, $L^\star_X$ is a linear operator. The operator $\Gamma$ and the matrix $\mathbf{M}$ define geometries in $\HH$ and $\RR^N$, respectively, through
\begin{equation}
\inHG{f}{g} = \inH{f}{\Gamma g},\quad f, g \in \HH, \quad\text{and}\quad \inRM{u}{v} = u^\top \mathbf{M} v,\quad u, v \in \RR^N.
\end{equation}
We denote by $\normHG{\cdot}$ and $\normRM{\cdot}$ their associated norms.
Using the definition of adjoint operators $L_X$ and $L^\star_X$, we have that
\begin{equation}
    \inRM{L_X(f)}{u} = \inHG{f}{L^\star_X(u)}, \quad\text{for all}\quad f \in \HH, u \in \RR^N.
\end{equation}
These relationships can be expressed as a duality diagram, see Figure~\ref{fig:duality_diagram}. The triplet $(X, \Gamma, \mathbf{M})$ defines a (multivariate) functional data analysis framework. One consequence of this transition between spaces is that the eigencomponents of $\mathcal{X}$ can be estimated equivalently using either the covariance operator $\Gamma$ or the Gram matrix $\mathbf{M}$. The relationship between the eigencomponents of $\Gamma$ and the eigencomponents of $\mathbf{M}$ are derived in Section~\ref{sec:functional_principal_components_analysis}.

\begin{remark}
   In general, in order to avoid confusion, inner-products and norms in function space, $\HH$ or $\sLp{\TT{}}$, will be refered with angle brackets, $\inLp{\cdot}{\cdot}$, while inner-products and norms in coordinate space, $\RR^N$,  will be refered with round brackets, $\inR{\cdot}{\cdot}$.
\end{remark}

\begin{remark}\label{rem:rhks}
We present here the duality diagram for the linear integral operator $\Gamma$ with kernel $C$. It is however possible to define duality diagrams for more general linear integral operators defined with a continuous symmetric positive definite function as kernel (see \cite{gonzalezRepresentingFunctionalData2010} and \cite{wongNonparametricOperatorregularizedCovariance2019a} for discussions on possible integral operators to represent univariate functional data).
\end{remark}

\subsection{Cloud of individuals} 
\label{sub:cloud_of_individuals}

Given an element $f \in \HH$, let $\{f^{(p)}(t_p),\,t_p \in \TT{p},\,p = 1, \dots, P\}$ be the features set of the element. We identify this set as the point $\pobs{M}_f$ in the space $\HH$. The space $\HH$ is referred to as the observation space. The cloud of points that represents the set of observations $\mathcal{X}$ in $\HH$ is denoted by $\CP$. Let $\Gmu$ be the centre of gravity of the cloud $\CP$. In the space $\HH$, its coordinates are given by $\{\mup{p}(t_p),\,t_p \in \TT{p},\,p = 1, \dots, P\}$. If the features are centered, the origin $\OH$ of the axes in $\HH$ coincides with $\Gmu$.

\begin{figure}
    \centering
    \includegraphics[scale=1.2]{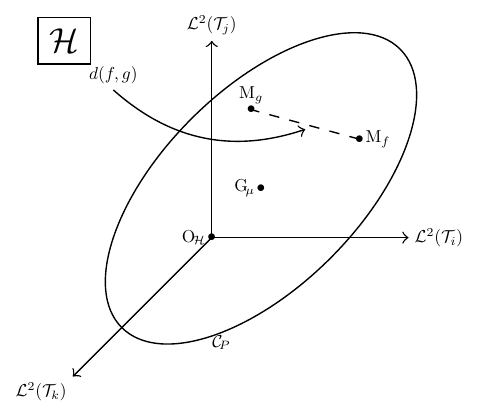}
    \includegraphics[scale=1.2]{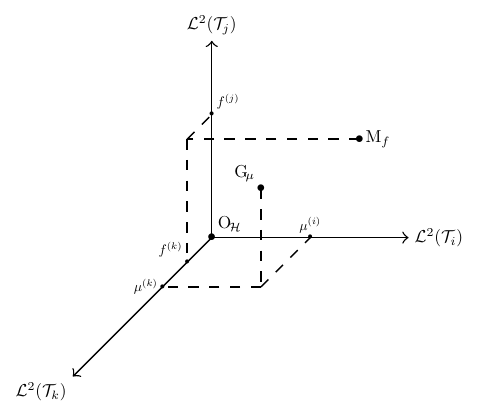}
    \caption{Left: Cloud of observations. Right: Projection of the points onto the elements of $\HH$. The observation $f$ (resp. $g$) is identified by the point $\pobs{M}_f$ (resp. $\pobs{M}_g$) in the cloud $\CP$. The point $\Gmu$ is the center of gravity of $\CP$ and the point $\OH$ is the origin of the space $\HH$.}
    \label{fig:cloud_obs}
\end{figure}

Let $f$ and $g$ be two elements in $\HH$ and denote by $\pobs{M}_f$ and $\pobs{M}_g$ their associated points in $\CP$ (see Figure~\ref{fig:cloud_obs}). The most natural distance between these observations is based on the usual inner product in $\HH$, $\inH{\cdot}{\cdot}$, and is defined as
\begin{equation}\label{eq:distance_obs}
    d^2(\pobs{M}_f, \pobs{M}_g) = \normH{f - g}^2 = \sum_{p = 1}^P \int_{\TT{p}}\left\{\fp(t_p) - \gp(t_p)\right\}^2 \dd t_p.
\end{equation}
This distance measures how different the observations are, and thus gives one characterization of the shape of the cloud $\CP$. Another description of this shape is to consider the distance between each observation and $\Gmu$, the center of the cloud. Let $f$ be an element of $\HH$, associated to the point $\pobs{M}_f$, and $\mu$ the element of $\HH$ related to $\Gmu$, the distance between $f$ and $\mu$ is given by
\begin{equation}\label{eq:distance_center}
    d^2(\pobs{M}_f, \Gmu) = \normH{f - \mu}^2 = \sum_{p = 1}^P \int_{\TT{p}}\left\{\fp(t_p) - \mup{p}(t_p)\right\}^2 \dd t_p.
\end{equation}
Given the set $\XX$, the total inertia of $\CP$, with respect to $\Gmu$ and the distance $d$, is given by
\begin{equation}\label{eq:inertia}
    \sum_{n = 1}^N \pi_n d^2(\pobs{M}_n, \Gmu) = \frac{1}{2}\sum_{i = 1}^N \sum_{j = 1}^N \pi_i \pi_j d^2(\pobs{M}_i, \pobs{M}_j) = \sum_{p = 1}^P \int_{\TT{p}}\Var{\Xp{p}(t_p)} \dd t_p.
\end{equation}
The duality diagram, however, allows us to define another suitable distance to characterize the shape of the cloud $\CP$. We thus define
\begin{equation}
    d^2_\Gamma(\pobs{M}_f, \pobs{M}_g) = \normHG{f - g}^2.
\end{equation}
The utilization of the distance measure $d_\Gamma$, which accounts for the variability among all the features within the functional data, corresponds to a Mahalanobis-type distance framework for multivariate functional data \cite[see][]{berrenderoMahalanobisDistanceFunctional2020,martinoKmeansProcedureBased2019}.
Given the set $\XX$, the total inertia of $\CP$, with respect to $\Gmu$ and the distance $d_\Gamma$, is given by
\begin{equation}\label{eq:inertia_CP}
    \sum_{n = 1}^N \pi_n d_\Gamma^2(\pobs{M}_n, \Gmu) = \frac{1}{2}\sum_{i = 1}^N \sum_{j = 1}^N \pi_i \pi_j d_\Gamma^2(\pobs{M}_i, \pobs{M}_j) = \sum_{p = 1}^P \int_{\TT{p}} \normH{C_{p \cdot}(t_p, \cdot)}^2 \dd t_p.
\end{equation}
The derivation of these equalities are given in Appendix \ref{sec:derivation_of_the_inertia_of_the_clouds}.

\begin{remark}
    These results have the same interpretation as for multivariate scalar data. This is also the multivariate analogue of the relation between variance and sum of squared differences known for univariate functional data. If the features are reduced beforehand, such that $\int_{\TT{p}} \Var\Xp{p} (t_p) \dd t_p = 1$ for the distance $d$ or $\int_{\TT{p}} \normH{C_{p \cdot}(t_p, \cdot)}^2 \dd t_p = 1$ for the distance $d_\Gamma$, the total inertia of the cloud $\CP$ is equal to the number of components $P$. We are, in general, not interested by the total inertia but how this variance is spread among the features.
\end{remark}


\subsection{Cloud of features} 
\label{sub:cloud_of_features}

Given an element $f \in \HH$, let $L_X(f) = \{\sqrt{\pi_n}\inH{X_n - \mu}{f},\,n = 1, \dots, N\}$ be the set of projections of $f$ onto the centered observations. We identify this set as the point $\pfea{M}_f$ in the space $\RR^N$. The space $\RR^N$ is referred to as the features' space. The cloud of points that represents the set of observations in $\RR^N$ is denoted by $\CN$. Let $\Gfea$ be the centre of gravity of the cloud $\CN$. In the space $\RR^N$, its coordinates are given by $L_X(\mu) = \{\sqrt{\pi_n}\inH{X_n - \mu}{\mu},\,n = 1, \dots, N\}$. If the data are centered, the origin $\OG$ of the axes in $\RR^N$ coincides with $\Gfea$.

\begin{figure}
    \centering
    \includegraphics[scale=1.2]{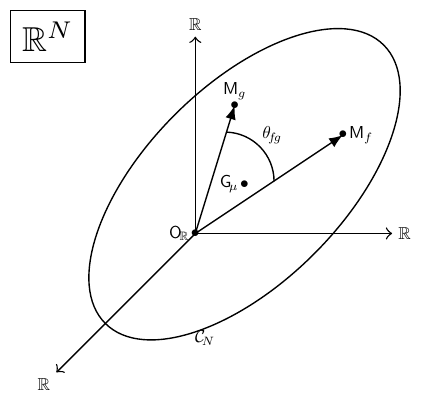}
    \includegraphics[scale=1.2]{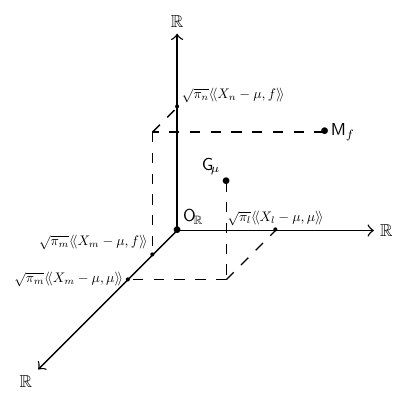}
    \caption{Left: Cloud of features. Right: Projection of the points on the elements of $\RR^N$. The observation $f$ (resp. $g$) is identified by the point $\pfea{M}_f$ (resp. $\pfea{M}_g$) in the cloud $\CN$. The point $\Gfea$ is the center of gravity of $\CN$ and the point $\OG$ is the origin of the space $\RR^N$.}
    \label{fig:cloud_features}
\end{figure}

We consider the usual inner-product in $\RR^N$, such that for all $u, v \in \RR^N, \inR{u}{v} = u^\top v$, associated with the norm $\normR{\cdot}$. Let $f$ and $g$ be two elements in $\HH$ and denote by $\pfea{M}_f$ and $\pfea{M}_g$ their associated points in $\CN$ (see Figure~\ref{fig:cloud_features}). The distance between $\pfea{M}_f$ and $\pfea{M}_g$ is thus defined as
\begin{equation*}
\mathsf{d}^2(\pfea{M}_f, \pfea{M}_g) = \normR{L_X(f) - L_X(g)}^2 = \sum_{n = 1}^N \pi_n \inH{X_n - \mu}{f - g}^2.
\end{equation*}
Similarly to the cloud of individuals, this distance characterizes the shape of the cloud $\CN$ and we also have access to this characterization through the distance with the center of gravity $\Gfea$. Let $f$ be an element of $\HH$, associated to the point $\pfea{M}_f$, and $\mu$ the element of $\HH$ related to $\Gfea$, the distance between $\pfea{M}_f$ and $\mu$ is given by
\begin{equation*}
\mathsf{d}^2(\pfea{M}_f, \Gfea) = \normR{L_X(f) - L_X(\mu)}^2 = \sum_{n = 1}^N \pi_n \inH{X_n - \mu}{f - \mu}^2.
\end{equation*}
Given the set $\XX$, the total inertia of $\CN$, with respect to $\Gfea$ and the distance $\mathsf{d}$, is given by
\begin{equation}\label{eq:inertia_CN}
    \sum_{n = 1}^N \pi_n \mathsf{d}^2(\pfea{M}_n, \Gfea) = \frac{1}{2}\sum_{i = 1}^N \sum_{j = 1}^N \pi_i \pi_j \mathsf{d}^2(\pfea{M}_i, \pfea{M}_j) = \sum_{p = 1}^P \int_{\TT{p}} \normH{C_{p \cdot}(t_p, \cdot)}^2 \dd t_p.
\end{equation}
Using the distances induced by the duality diagram, the total inertia of the cloud $\CN$ is thus equal to the total inertia of the cloud $\CP$. This property highlights the duality between the spaces $\HH$ and $\RR^N$. To further emphasize this duality, the cosine of the angle $\theta_{fg}$ formed by the two points $\pfea{M}_f$ and $\pfea{M}_g$ is equal to their correlation coefficient and can be written
\begin{equation}
    \cos(\theta_{fg}) = \frac{\inR{L_X(f)}{L_X(g)}}{\normR{L_X(f)}\normR{L_X(g)}}
    = \frac{\inHG{f}{g}}{\normHG{f}\normHG{g}}.
\end{equation}
The derivation of these equalities are given in Appendix \ref{sec:derivation_of_the_inertia_of_the_clouds}.

\begin{remark}
   Although each axis of the space does not directly represent the features, but rather the projection of an element of $\HH$ onto the elements of the set $\XX$, we refer to this space as the features' space. We use this terminology to highlight the similarity between multivariate functional data analysis and traditional multivariate data analysis, as well as to emphasize the dimensionality of this space.
\end{remark}

\subsection{On centering and reducing} 
\label{sub:on_centering_and_reducing}

For conducting an MFPCA, the features are usually assumed centred \citep{happMultivariateFunctionalPrincipal2018a}. \cite{protheroNewPerspectivesCentering2023} give a complete overview of centering in the context of FDA. Here, we comment on the geometric interpretation of centering in this context and compare with the multivariate scalar case. We focus on the usual centering in FDA, namely $\Xnp(t_p) - \mu^{(p)}(t_p),~t_p \in \TT{p}$ (referred to \emph{object centering} in \cite{protheroNewPerspectivesCentering2023}).
The geometric interpretation of object centering is the same if we refer to the observation space $\HH$ or the feature space $\RR^N$. Within the space $\HH$ (resp. $\RR^N$), centering is interpreted as translating the centre of gravity of the clouds, $\Gmu$ (resp. $\Gfea$), to the origin point $\OH$ (resp. $\OG$) of the space $\HH$ (resp. $\RR^N$). This transformation, being a translation, does not change the shape of the cloud $\CP$ (resp. $\CN$). The interpretation is the same as for the centering in the multivariate scalar data context within their observation space.

Concerning the standardization of the data, there are two main proposals in the literature. \cite{happMultivariateFunctionalPrincipal2018a} propose to consider the weights
\begin{equation}
w_p = \left(\int_{\TT{p}} \Var X^{(p)}(t_p) \dd t_p\right)^{-1}.
\end{equation}
This standardization is coherent with the derivation of the total inertia of the observation space using the usual distance in $\HH$. \cite{chiouMultivariateFunctionalPrincipal2014} and \cite{jacquesModelbasedClusteringMultivariate2014a} propose the weighting function
\begin{equation}
w_p(t_p) = \left(\Var X^{(p)}(t_p)\right)^{-1}, \quad t_p \in \TT{p}.
\end{equation}
This corresponds to a standardization of the curves by the standard deviation of the component at each sampling point. The standard deviation curve is estimated as the square root of the diagonal of the covariance function estimates, obtained using a local linear smoother of the pooled data. For each functional feature $p$, this standardization mimics the standardization used for principal components analysis if the number of (scalar) features is infinite.
Considering the duality diagram and the total inertia of the clouds with respect to the distance $d_\Gamma$, we propose to use the weights
\begin{equation}
    w_p = \left(\int_{\TT{p}} \normH{C_{p \cdot}(t_p, \cdot)}^2 \dd t_p\right)^{-1}.
\end{equation}
The total inertia of $\CP$ (and $\CN$) will thus be equal to the number of components.
In every case, we may consider the rescaled elements $\widetilde{X}^{(p)} = w_p^{1/2}\{X^{(p)} - \mup{p}\}$ in place of the elements $X^{(p)}$.



\section{Multivariate functional principal components analysis} 
\label{sec:functional_principal_components_analysis}

Assuming that the covariance operator $\Gamma$ is a compact positive operator on $\HH$ and using the results in \cite{happMultivariateFunctionalPrincipal2018a}, and the theory of Hilbert-Schmidt operators, e.g., \cite{reedMethodsModernMathematical1980}, there exists a complete orthonormal basis 
$\Phi = \{\phi_k\}_{k \geq 1} \subset \HH$ associated to a set of real numbers $\{\lambda_k\}_{k \geq 1}$ such that $\lambda_1 \geq \lambda_2 \geq \dots \geq 0$ that satisfy
\begin{equation}\label{eq:eigendecomposition}
    \Gamma \phi_k = \lambda_k \phi_k, \quad\text{and}\quad \lambda_k \longrightarrow 0 \quad\text{as}\quad k \longrightarrow \infty.
\end{equation}
The set $\{\lambda_k\}_{k \geq 1}$ contains the eigenvalues of the covariance operator $\Gamma$ and $\Phi$ contains the associated eigenfunctions. Using the multivariate Karhunen-Loève theorem \citep{happMultivariateFunctionalPrincipal2018a}, we obtain the decomposition
\begin{equation}\label{eq:kl_multi}
    X(\pointt) = \mu(\pointt) + \sum_{k = 1}^\infty \mathfrak{c}_k \phi_k(\pointt), \quad \pointt \in \TT{}
\end{equation}
where $\mathfrak{c}_{k} = \inH{X - \mu}{\phi_k}$ are the projections of the centered curves onto the eigenfunctions. We have that $\EE(\mathfrak{c}_{k}) = 0$, $\EE(\mathfrak{c}_{k}^2) = \lambda_k$ and $\EE(\mathfrak{c}_{k}\mathfrak{c}_{k^\prime}) = 0$ for $k \neq k^\prime$. Note that the coefficients $\mathfrak{c}_k$ are scalar random variables while the multivariate functions $\phi_k$ are vectors of functions. Let us call $\Phi$ the multivariate functional principal component analysis basis. In practice, we use a truncated version of the Karhunen-Loève expansion \eqref{eq:kl_multi} as the eigenvalues $\lambda_k$, and hence the contribution of $\mathfrak{c}_k$ to \eqref{eq:kl_multi}, becomes negligible as $k$ goes to infinity. Let
\begin{equation}\label{eq:kl_multi_trunc}
    X_{\lceil K \rceil}(\pointt) = \mu(\pointt) + \sum_{k = 1}^K \mathfrak{c}_k \phi_k(\pointt), \quad \pointt \in \TT{}, \quad K \geq 1,
\end{equation}
be the truncated Karhunen-Loève expansion of the process $X$ and
\begin{equation}\label{eq:kl_multi_trunc_comp}
    X_{\lceil K_p \rceil}^{(p)}(t_p) = \mup{p}(t_p) + \sum_{k = 1}^{K_p} \mathfrak{c}^{(p)}_k \varphi_k^{(p)}(t_p), \quad t_p \in \TT{p}, \quad K_p \geq 1, \quad 1 \leq p \leq P,
\end{equation}
be the truncated Karhunen-Loève expansion of the $p$th feature of the process $X$. For each $p$, the function $\mup{p}$ is the $p$th feature of the multivariate mean function $\mu$ and the set $\{\varphi^{(p)}_k\}_{1 \leq k \leq K_p}$ is a basis of univariate functions in $\sLp{\TT{p}}$, whose elements are not the components of the multivariate functions $\phi_k$. In~\eqref{eq:kl_multi_trunc_comp}, the coefficients $\mathfrak{c}^{(p)}_k$ are the projection of the centered curve $\Xp{p}$ onto the eigenfunctions $\varphi_k^{(p)}$ and are not (directly) related to the coefficients $\mathfrak{c}_k$ in~\eqref{eq:kl_multi_trunc}.

\subsection{Diagonalization of the covariance operator} 
\label{sub:by_diagonalization_of_the_covariance_operator}

The estimation of the eigencomponents of the covariance $\Gamma$ by its diagonalization is derived in \cite{happMultivariateFunctionalPrincipal2018a} for a general class of multivariate functional data defined on different dimensional domains. They give a direct relationship between the truncated representation \eqref{eq:kl_multi_trunc_comp} of the single elements $X^{(p)}$ and the truncated representation \eqref{eq:kl_multi_trunc} of the multivariate functional data $X$.

We recall here how to estimate the eigencomponents. Following \citet[Prop.~5]{happMultivariateFunctionalPrincipal2018a}, the multivariate components for $X$ are estimated by a weighted combination of the univariate components computed from each $X^{(p)}$. First, we perform a univariate FPCA on each of the features of $X$ separately. For a feature $X^{(p)}$, the eigenfunctions and eigenvectors are computed using a matrix decomposition of the estimated covariance $C_{p, p}$ from \eqref{eq:cov_estimation}. This results in a set of eigenfunctions $\{\varphi_k^{(p)}\}_{1 \leq k \leq K_p}$ associated with a set of eigenvalues $\{\lambda_k^{(p)}\}_{1 \leq k \leq K_p}$ for a given truncation integer $K_p$. Then, the univariate scores for a realization $\Xnp$ of $X^{(p)}$ are given by $\mathbf{c}_{nk}^{(p)} = \inLp{X_n^{(p)}}{\varphi_k^{(p)}}, ~1 \leq k \leq K_p$. These scores might be estimated by numerical integration for example. Considering $K_+ \coloneqq \sum_{p = 1}^P K_p$, we then define the matrix $\mathcal{Z} \in \mathbb{R}^{N \times K_+}$, where on each row we concatenate the scores obtained for the $P$ features of the $n$th observation: 
$(\mathbf{c}_{n1}^{(1)}, \ldots, \mathbf{c}_{nK_1}^{(1)}, \ldots, \mathbf{c}_{n1}^{(P)}, \ldots, \mathbf{c}_{nK_P}^{(P)})$. An estimation of the covariance of the matrix $\mathcal{Z}$ is given by $\mathbf{Z} = (N - 1)^{-1}\mathcal{Z}^\top\mathcal{Z}$. An eigenanalysis of the matrix $\mathbf{Z}$ is carried out to estimate the eigenvectors $\boldsymbol{v}_k$ and eigenvalues $\lambda_k$. Finally, the multivariate eigenfunctions are estimated as a linear combination of the univariate eigenfunctions using
\begin{equation*}
\phi_k^{(p)}(t_p) = \sum_{l = 1}^{K_p}[\boldsymbol{v}_k]_{l}^{(p)}\varphi_{l}^{(p)}(t_p),\quad t_p \in \TT{p},\quad 1 \leq k \leq K_+,\quad 1 \leq p \leq P,
\end{equation*}
where $[\boldsymbol{v}_k]^{(p)}_{l}$ denotes the $l$th entry of the $p$th block of the vector $\boldsymbol{v}_k$. The multivariate scores are estimated as
$$\mathfrak{c}_{nk} = \mathcal{Z}_{{n,\cdot}}\boldsymbol{v}_k, \quad 1 \leq n \leq N, \quad 1 \leq k \leq K_+,$$
where $\mathcal{Z}_{{n,\cdot}}$ is the $n$th row of the matrix $\mathcal{Z}$.
We refer the reader to \cite{happMultivariateFunctionalPrincipal2018a} for the derivation of the eigencomponents of the covariance operator if the curves are expanded in a general basis of functions.


\subsection{Diagonalization of the inner product matrix} 
\label{sub:by_diagonalization_of_the_inner_product_matrix}

We use the duality relation between row and column spaces of a data matrix to estimate the eigencomponents of the covariance operator. Consider the inner-product matrix $\mathbf{M}$, with entries defined in~\eqref{eq:gram_mat} and assuming that all observations are equally weighted, i.e., for all $n = 1, \dots, N$, $\pi_n = 1/N$.
Let $\{l_k\}_{1 \leq k \leq N}$ such that $l_1 \geq \dots \geq l_N \geq 0$ be the set of eigenvalues and $\{\boldsymbol{u}_k\}_{1 \leq k \leq N}$ be the set of eigenvectors of the matrix $\mathbf{M}$. The relationship between all nonzero eigenvalues of the covariance operator $\Gamma$ and the eigenvalues of $\mathbf{M}$ is given by
\begin{equation}\label{eq:eigenvalues_relation_p}
    \lambda_k = l_k, \quad k = 1, 2, \dots, N,
\end{equation}
while the relationship between the multivariate eigenfunctions of the covariance operator $\Gamma$ and the orthonormal eigenvectors of $M$ is given by
\begin{equation}\label{eq:eigenfunction_relation_p}
    \phi_k(\pointt) = \frac{1}{\sqrt{N l_k}}\sum_{n = 1}^N [\boldsymbol{u}_{k}]_n\left\{X_n(\pointt) - \mu(\pointt)\right\}, \quad \pointt \in \TT{}, \quad k = 1, 2, \dots, N, 
\end{equation}
where $[\boldsymbol{u}_{k}]_n$ is the $n$th entry of the vector $\boldsymbol{u}_k$. The scores are then computed as the inner-product between the multivariate curves and the multivariate eigenfunctions and are given by
\begin{equation}\label{eq:scores_relation_p}
    \mathfrak{c}_{nk} = \sqrt{N l_k}[\boldsymbol{u}_{k}]_n, \quad n = 1, 2, \dots, N, \quad k = 1, 2, \dots, N. 
\end{equation}
The derivations of these equalities are given in Appendix~\ref{sec:derivation_of_the_eigencomponents} in a slighty more general framework where the observation weights $\pi_n$ are not equal. These results can be extended in a natural way if all the curves are expanded in a general basis of functions, see Section \ref{sub:with_a_basis_expansion} in the Supplementary Material.


\subsection{Computational complexity} 
\label{sub:computational_complexity}

We describe the time complexity for the computation of the MFPCA algorithm using the covariance operator and the Gram matrix. Considering the observation of $N$ curves with $P$ features, we assume that all observations of feature $p$ are sampled on a common grid of $M_p$ points. For $a \in \NN$, let $M^a = \sum_{p} M_p^a$. Let $K$ be the number of multivariate eigenfunctions to estimate. For the estimation of the eigencomponents using the covariance operator, we have $K \leq K_+$. While $K$ has the same interpretation for both the eigendecomposition of the covariance operator and the eigendecomposition of the inner product matrix, in the latter case, it is not computed as the summation over the univariate elements, but rather as the number of components needed to achieve a certain amount of variance explained. Here, we also assume that the curves are perfectly observed, and thus no smoothing step is included in the expression of the time complexity. Note that the smoothing step will often have the same impact on complexity between the approaches as the smoothing is a preprocessing step.

To estimate the time complexity, we count the number of elementary operations performed, considering a fixed execution time for each. Worst-case time complexity is considered. We first give the time complexity for the estimation of the eigencomponents using the covariance operator by explaining the time complexity of each individual step (see \cite{happMultivariateFunctionalPrincipal2018a} and Section~\ref{sub:by_diagonalization_of_the_covariance_operator}). For each feature $p$, the time complexity of the estimation of the covariance matrix is $\bigO(NM_p^2)$, of the eigendecomposition of the matrix is $\bigO(M_p^3)$ and of the univariate score is $\bigO(NM_pK_p)$. Therefore, the total time complexity is the sum over the $p$ univariate time complexities. The covariance matrix $\mathbf{Z}$ of the stacked univariate scores $\mathcal{Z}$ is then computed with a time complexity of $\bigO(NK_+^2)$, because the dimension of the matrix $\mathcal{Z}$ is $N \times K_+$. The eigendecomposition of the matrix $\mathbf{Z}$ has a time complexity of $\bigO(K_+^3)$. The final step is to compute the multivariate eigenfunctions and scores. For the estimation of the $K \leq K_+$ multivariate eigenfunctions, the time complexity is $\bigO(K\sum_{p} M_pK_p)$ and for the estimation of the scores, the time complexity is $\bigO(NK^2)$. Gathering all the results, the final complexity of the estimation of the eigencomponents using the eigendecomposition of the covariance operator is
\begin{equation}\label{eq:time_compl_cov}
    \bigO\left(NM^2 + M^3 + N\sum_{p = 1}^P M_pK_p + NK_+^2 + K_+^3 + K\sum_{p = 1}^P M_pK_p + NK^2\right).
\end{equation}
We now consider the time complexity of the estimation of the eigencomponents using the eigendecomposition of the inner product matrix (see Section~\ref{sub:by_diagonalization_of_the_inner_product_matrix}). The inner product between two curves can be estimated in $\bigO(M^1)$. Since there are $N^2$ terms in the matrix, the time complexity for the computation of the inner product matrix is then $\bigO(N^2M^1)$. The eigendecomposition of this matrix has a time complexity of $\bigO(N^3)$. For the multivariate eigenfunctions, the time complexity is $\bigO(KNP)$ and is $\bigO(KN)$ for the multivariate scores. Gathering all the results, the final complexity of the estimation of eigencomponents using the eigendecomposition of the inner product matrix is
\begin{equation}\label{eq:time_compl_in_prod}
    \bigO\left(N^2M^1 + N^3 + KNP + KN\right).
\end{equation}

The number of components $K$ to estimate is usually small compared to the number of curves $N$ or to the total number of sampling points $M^1$. Both time complexities can then be reduced to $\mathcal{O}(NM^2 + M^3)$ for the diagonalization of the covariance operator and to $\mathcal{O}(N^2M^1 + N^3)$ using the Gram matrix. If the number of observations is large compared to the total number of sampling points, it thus seems preferable to use the covariance operator to estimate the eigencomponents, while if the total number of sampling points is large compare to the number of observations, the use of the Gram matrix seems better. Note that the number of features $P$ does not have much impact on the computational complexity, in the sense that the important part is the total number of sampling points. One component with $1000$ sampling points will have the same computational complexity as $100$ components with $10$ sampling points. These results are confirmed in the simulation (see Section~\ref{sub:simulation_results}).

\begin{remark}
We can use singular values decomposition (SVD) in both cases to make the algorithm faster as it allows to compute only the first $K$ eigenfunctions. In practice, this might be important as the maximum number of non-zero eigenvalues is the minimum between the number of observations and the number of sampling points.
\end{remark}



\section{Empirical analysis} 
\label{sec:empirical_analysis}

Using simulated data, we compare the estimation of the eigencomponents and the reconstruction of the curves using the diagonalization of the covariance operator, (a) based on univariate FPCA as well as (b) P-splines basis expansion, and (c) the diagonalization of the Gram matrix. These methods will be refered as (a) \texttt{(Tensor) PCA}, (b) \texttt{2D/1D B-Splines} and (c) \texttt{Gram} respectively. For the diagonalization of the covariance operator, we consider the methodology of \cite{happMultivariateFunctionalPrincipal2018a}. In the case of MFPCA based on the expansion of each univariate feature into univariate principal components (\texttt{(Tensor) PCA}), the eigencomposition of the 1-dimensional curves is performed using univariate FPCA and the eigendecomposition of the 2-dimensional curves is calculated using the Functional Canonical Polyadic-Tensor Power Algorithm (FCP-TPA) for regularized tensor decomposition \citep{allenMultiwayFunctionalPrincipal2013a}. In the case of MFPCA based on P-splines basis expansion (\texttt{2D/1D B-Splines}), data are expanded in B-splines basis with suitable smoothness penalty \citep{eilersFlexibleSmoothingBsplines1996}.

The results of the simulation are compared using computation times (CT), the integrated squared error (ISE) for the multivariate eigenfunctions, the relative squared error (RSE) for the eigenvalues and the mean relative squared error (MRSE) for the reconstructed data. Let $\phi_k$ be the true eigenfunction and $\widehat{\phi}_k$ the estimated eigenfunction defined on $\TT{}$. We then define the ISE as 
\begin{equation}\label{eq:ise_eigenfunctions}
    \text{ISE}(\phi_k, \widehat{\phi}_k) = \normH{\phi_k - \widehat{\phi}_k}^2 = \sum_{p = 1}^P \int_{\TT{p}} \{\phi^{(p)}_k(t_p) - \widehat{\phi}^{(p)}_k(t_p)\}^2 \dd t_p, \quad k = 1, \dots, K.
\end{equation}
Let $\lambda = \{\lambda_1, \dots, \lambda_K\}$ be the set of true eigenvalues and $\widehat{\lambda} = \{\widehat{\lambda}_1, \dots, \widehat{\lambda}_K\}$ be the set of estimated eigenvalues. We then define the RSE as 
\begin{equation}\label{eq:mse_eigenvalues}
    \text{RSE}(\lambda_k, \widehat{\lambda}_k) = \left(\lambda_k - \widehat{\lambda}_k\right)^2 / \lambda_k^2, \quad k = 1, \dots, K.
\end{equation}
Let $\mathcal{X}$ be the set of true data and $\widehat{\mathcal{X}}$ be the set of reconstructed data. We define the MISE of the reconstructed data as
\begin{equation}\label{eq:mise_reconstructed_data}
    \text{MRSE}(\mathcal{X}, \widehat{\mathcal{X}}) = \frac{1}{N}\sum_{n = 1}^N \normH{X_n - \widehat{X}_n}^2 = \frac{1}{N}\sum_{n = 1}^N \sum_{p = 1}^P \int_{\TT{p}} \left\{\Xnp(t_p) - \hatXnp{p}(t_p) \right\}^2 \dd t_p.
\end{equation}
Each integral is approximated by the trapezoidal rule with an equidistant grid.

\subsection{Simulation experiments} 
\label{sub:simulation_experiments}

The simulation setting is based on the \emph{Setting 3} of the simulation in \cite{happMultivariateFunctionalPrincipal2018a}.
The data generating process is based on a truncated version of the Karhunen-Loève decomposition. We simulate multivariate functional data with $P = 2$ components. For the first component, we generate an orthonormal basis $\{\phi^{(1)}_k\}_{1 \leq k \leq K}$ of $\sLp{\TT{1}}$ on an interval $\TT{1} = [0, 1] \times [0, 0.5]$ as the tensor product of the first Fourier basis functions:
\begin{equation}
    \phi^{(1)}_k(s, t) = \psi_l(s) \otimes \psi_m(t), \quad s \in [0, 1] \text{ and } t \in [0, 0.5],\quad k = 1, \dots, K,
\end{equation}
where $\psi_l$ and $\psi_m$ are elements of the Fourier basis. For the second component, we generate an orthonormal Legendre basis $\{\phi^{(2)}_k\}_{1 \leq k \leq K}$ of $\sLp{\TT{2}}$ on an interval $\TT{2} = [-1, 1]$. To ensure orthonormality, the basis $\phi^{(1)}_k$ and $\phi^{(2)}_k$ are weighted by random factors $\alpha^{1/2}$ and $(1 - \alpha)^{1/2}$, respectively, where $\alpha \sim \mathcal{U}(0.2, 0.8)$.
Each curve is then simulated using the truncated multivariate Karhunen-Loève expansion \eqref{eq:kl_multi_trunc}:
\begin{equation}
    X(\pointt) = \sum_{k = 1}^K \mathfrak{c}_k \phi_k(\pointt), \quad \pointt \in \TT{} \coloneqq \TT{1} \times \TT{2},
\end{equation}
where $\phi_k = (\phi^{(1)}_k, \phi^{(2)}_k)^\top$ and the scores $\mathfrak{c}_k$ are sampled as random normal variables with mean $0$ and variance $\lambda_k$. The eigenvalues $\lambda_k$ are defined with an exponential decrease, $\lambda_k = \exp(-(k + 1)/2)$ We simulate, for each replication of the simulation, $N = 50, 100$ and $250$ observations. The first component is sampled on a regular grid of $M^{(1)} = 11 \times 11, 26 \times 26$ and $101 \times 51$ sampling points. The second component is sampled on a regular grid of $M^{(2)} = 21, 51$ and $201$ sampling points. We set $K = 25$.
We also consider data with measurement errors. In that case, we observe data from \eqref{eq:model_error}, where $\sigma^2 = (0.25, 0.25)$. Finally, we study sparse data as in \cite{happMultivariateFunctionalPrincipal2018a}, with medium ($50\%-70\%$ missing) and high ($90\%-95\%$ missing) sparsity in both components. For this setting, no noise is added to the data.

The \texttt{(Tensor) PCA} method is based on the univariate estimation of $K^{(1)} = 20$ eigenimages and $K^{(2)} = 15$ eigenfunctions. The estimation of the eigenimages is performed with the FCP-TPA where the smoothing parameters are chosen via cross-validation in $[10^{-5}, 10^5]$ for dense, sparse and noisy data. In the case of sparse data, we also linearly interpolate the two-dimensional observations to get data that are regularly sampled to run the FCP-TPA. The estimation of the eigenfunctions is done using the PACE algorithm \citep{yaoFunctionalDataAnalysis2005} with P-splines to smooth the mean and covariance functions. Penalties are chosen using cross-validation \citep{eilersFlexibleSmoothingBsplines1996}.

For the \texttt{2D/1D B-splines} method, $X^{(1)}$ is expanded in tensor products of $K^{(1)} = 13 \times 13$ B-splines and $X^{(2)}$ is expanded in $K^{(2)} = 13$ B-splines. For dense data, no penalty is involved, while for sparse and noisy data, penalties are chosen using cross-validation.

Finally, the \texttt{Gram} method is based on the observed data points in the dense case, on the linear interpolation of the data in the sparse case \citep{benkoCommonFunctionalPrincipal2009} and on the P-splines smoothing of the data in the noisy case.


\subsection{Simulation results} 
\label{sub:simulation_results}

We compared MFPCA results of the different methods in terms of their CT, estimation of eigenvalues (RSE), estimation of eigenfunctions (ISE), and reconstruction of curves (MRSE). We fix the number of retained components to be $12$ for each simulation scenario. Each experiment is repeated $200$ times. The results are presented below.

\begin{results}[Computational time.]
To compare the computational time of the different methods, we measured the time taken for each method to complete the MFPCA for each simulated dataset in the dense and noiseless case. Figure~\ref{fig:computation_time_mfd_1d} shows the kernel density estimates of the ratio of CT for each method across all sample sizes and number of sampling points. Except when we have many observations with few sampling points, the \texttt{Gram} method is faster than the other two methods. This result is coherent with the computational complexity derived in Section~\ref{sub:computational_complexity}.

\begin{figure}
     \centering
    \includegraphics[width=0.95\textwidth]{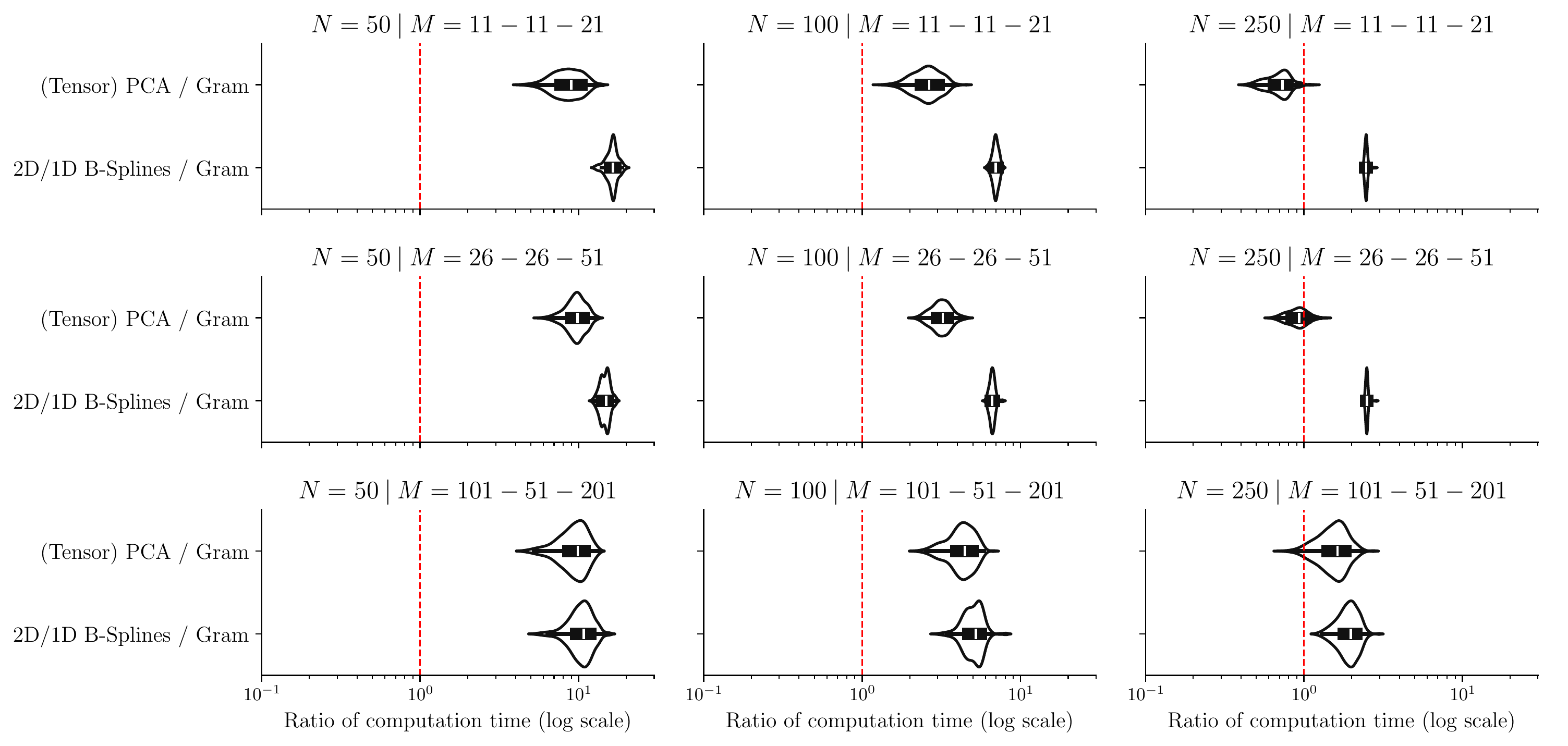}
    \caption{Ratio of computation time in the dense case between the \texttt{(Tensor) PCA} and \texttt{2D/1D B-Splines} methods and the \texttt{Gram} method. $N$ is the number of observations, $M$ is the number of sampling points per curve (the first two numbers are for the images and the last one is for the curves).}
    \label{fig:computation_time_mfd_1d}
\end{figure}

The shorter computational time of the diagonalization of the Gram matrix makes it a more efficient option for analyzing two and higher-dimensional functional datasets as the number of sampling points tend to grow faster in these situations. However, the computational time can still vary depending on the specific implementation of each method, the computational resources available, the complexity of the dataset (number of observations, number of sampling points, etc.) and whether a smoothing step needs to be included.
\end{results}

\begin{results}[Eigenvalue estimation.]
To compare the estimation of the eigenvalues between the different methods, we calculated the RSE \eqref{eq:mse_eigenvalues} between the estimated eigenvalues and the true eigenvalues for each simulated dataset and for the twelve estimated eigenvalues.
Figure~\ref{fig:logAE_mfd_1d} shows the boxplots of the RSE for each method across all sample sizes and number of sampling points. We found that the three methods behave similarly for the settings with a moderately large to a large number of sampling points. When we observe only a few sampling points, $M^{(1)} = 11 \times 11$ and $M^{(2)} = 21$, the quality of the estimation is approximately the same for the first three eigenvalues for all methods, but from the fourth eigenvalues, the \texttt{2D/1D B-splines} and the \texttt{Gram} methods give slightly better results than the \texttt{(Tensor) PCA} method. The results for the sparse and noisy cases are similar to the dense and noiseless case and are provided in the Appendix \ref{sub:simulation}.

\begin{figure}
    \centering
    \includegraphics[width=0.95\textwidth]{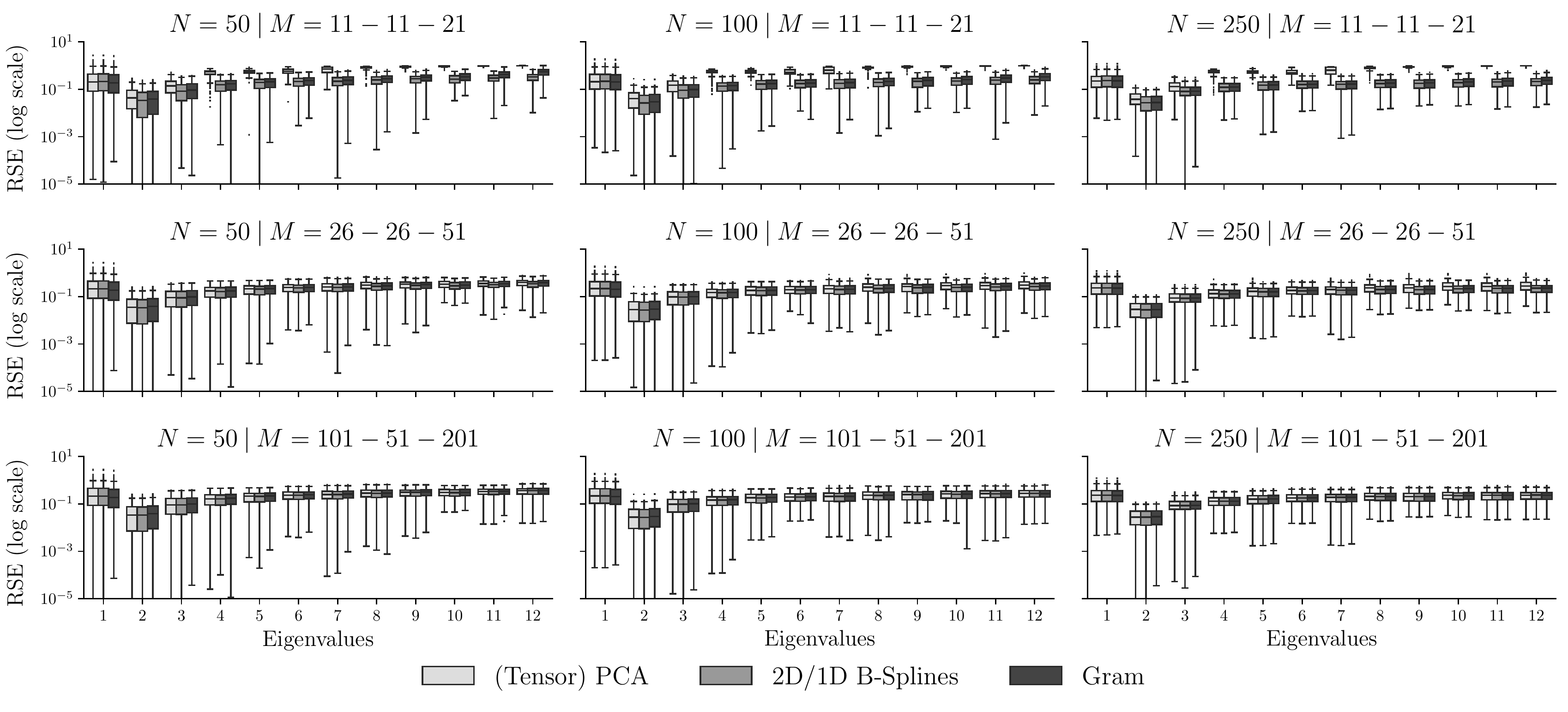}
    \caption{RSE for the estimated eigenvalues for each method in the dense case. $N$ is the number of observations, $M$ is the number of sampling points per curve (the first two numbers are for the images and the last one is for the curves).}
    \label{fig:logAE_mfd_1d}
\end{figure}
\end{results}

\begin{results}[Eigenfunction estimation.]
To compare the estimation of the eigenfunctions between the different methods, we calculated the ISE \eqref{eq:ise_eigenfunctions} between the estimated eigenfunctions and the true eigenfunctions for each simulated dataset and for the twelve estimated eigenfunctions. Figure~\ref{fig:ise_mfd_1d} shows the boxplots of the ISE for each method across all sample sizes and number of sampling points. The results are similar to those found when estimating the eigenvalues. When we observe only a few sampling points, the quality of the estimation of the eigenfunctions starts to decrease after the third component for the \texttt{(Tensor) PCA} method, while this is not the case for the other methods. The results for the sparse and noisy cases are similar to the dense and noiseless case and are provided in the Appendix \ref{sub:simulation}.
\begin{figure}
     \centering
    \includegraphics[width=0.95\textwidth]{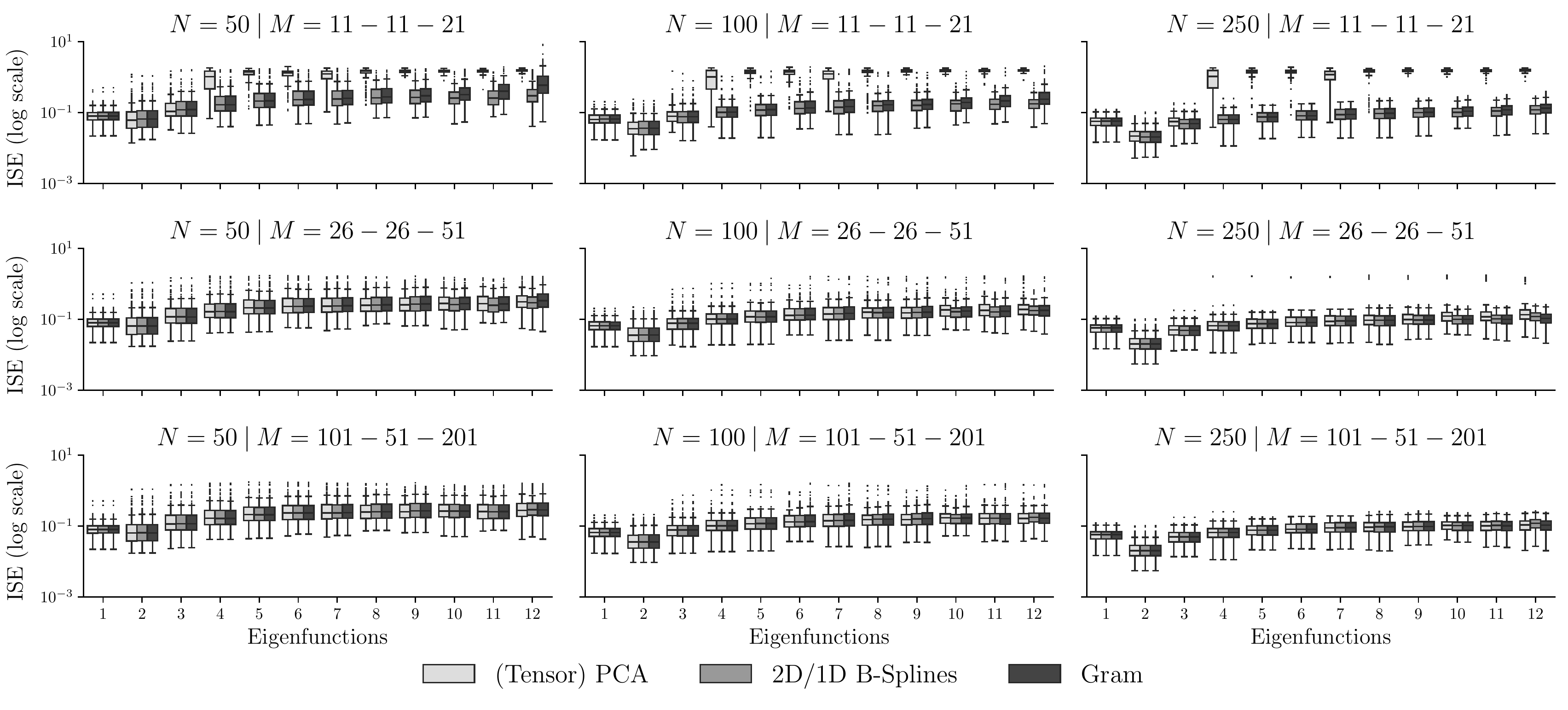}
    \caption{ISE for the estimated eigenfunctions for each method in the dense case. $N$ is the number of observations, $M$ is the number of sampling points per curve (the first two numbers are for the images and the last one is for the curves).}
    \label{fig:ise_mfd_1d}
\end{figure}
\end{results}

\begin{results}[Curve reconstruction.]
To compare the quality of the reconstruction of the curves between the different methods, we calculated the MRSE \eqref{eq:mise_reconstructed_data} between the reconstruction of the curves and the true curves for each simulated dataset. Figure~\ref{fig:mise_mfd_1d} shows the boxplots of the MRSE for each method across all sample sizes and number of sampling points. Except in the case of few sampling points where the \texttt{(Tensor) PCA} method does not perform well (because of the poor estimation of the eigenvalues and eigenfunctions), all methods give nearly the same results for each setting. The results for the sparse and noisy cases are similar and are provided in the Appendix \ref{sub:simulation}.
\begin{figure}
     \centering
     \includegraphics[width=0.95\textwidth]{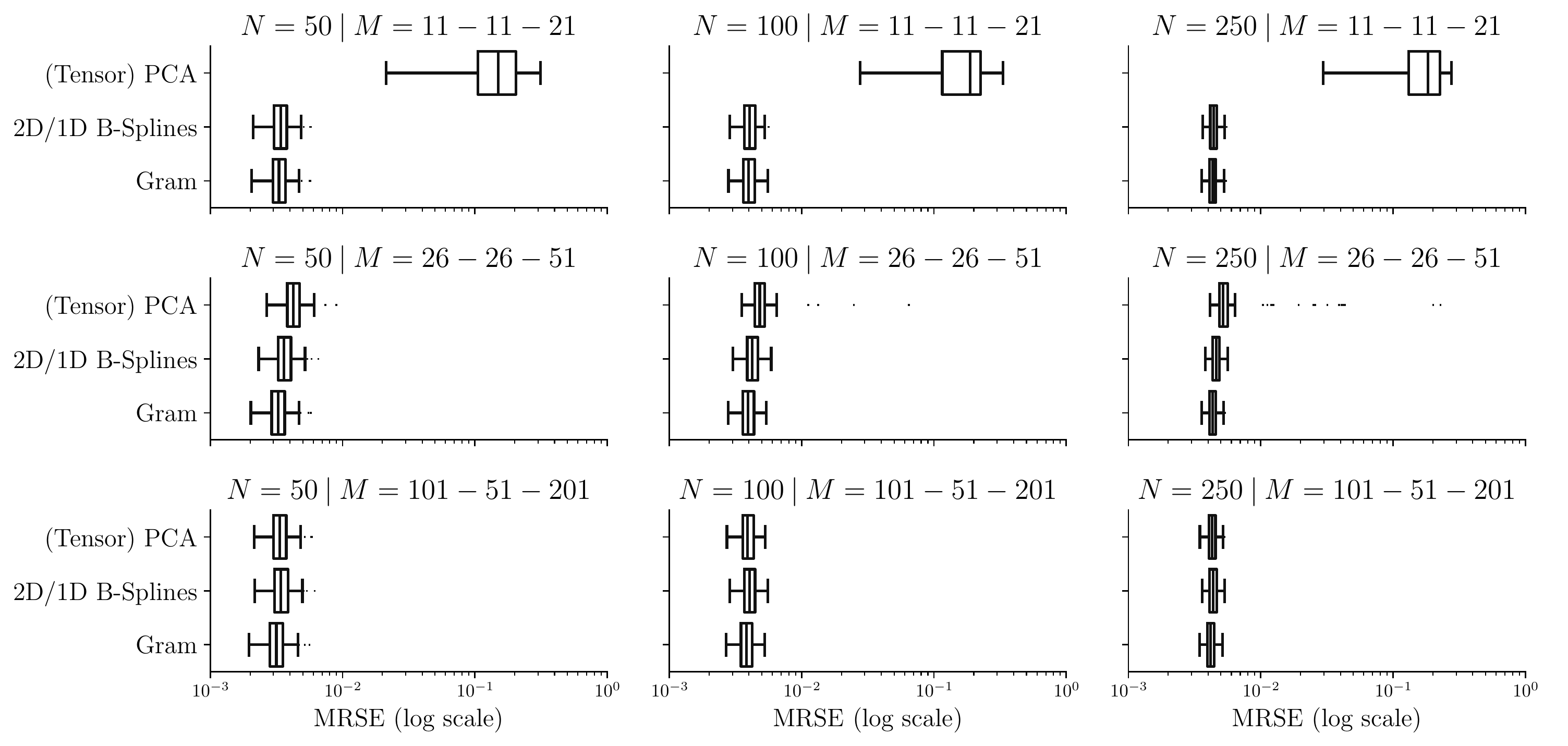}
    \caption{MRSE for the reconstructed curves for each method in the dense case. $N$ is the number of observations, $M$ is the number of sampling points per curve (the first two numbers are for the images and the last one is for the curves).}
    \label{fig:mise_mfd_1d}
\end{figure}
\end{results}

\section{Application} 
\label{sec:application}

\begin{figure}
    \centering
    \includegraphics[width=\textwidth]{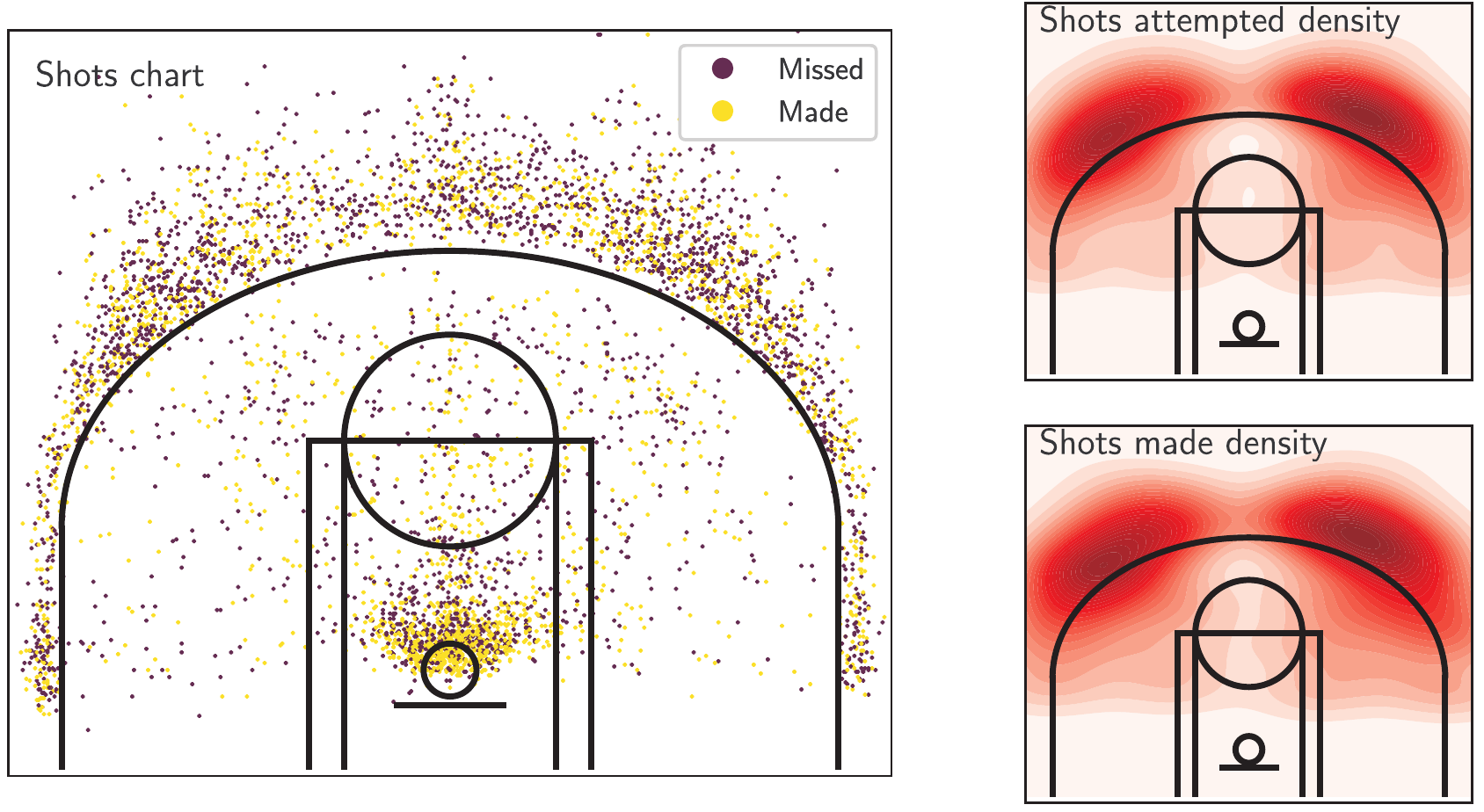}
    \caption{Made/Missed shots chart (left) and the estimated densities (right) for Stephen Curry.}
    \label{fig:shoots_make_miss}
\end{figure}

We apply our methodology utilizing data to the National Basketball Association (NBA). To access comprehensive game data, we utilize the Python package \texttt{nba\_api}\footnote{\url{https://github.com/swar/nba_api/}}, which provides access to the APIs of \url{nba.com}. The dataset encompasses shot location data from all NBA games spanning the seasons between $2018-2019$ and $2022-2023$. Filtering the dataset, we focus solely on players who have made more than $1000$ shots during this five-season period, resulting in a cohort of $131$ players. These players accounted for a total of $493723$ shots attempted, of which $234941$ ($47\%$) were successful. Subsequently, we exclude shots deemed impossible (e.g., out-of-bounds), leaving us with a dataset comprising $492621$ shots (see Figure \ref{fig:shoots_make_miss} for the shots chart of Stephen Curry). We remove all the shots that are close to the hoop (a square of $2.7 \times 2.8$ meters around the hoop), as these shots will be the most common and their pattern is not interesting from a shooting behavior perspective, and players that have made fewer than $100$ shots during the five-season period. This results in a cohort of $119$ players, with $186621$ attempted shots and $71893$ ($38.5\%$) made shots.

To analyze shooting behavior, we employ a 2-dimensional kernel density estimation for both the attempted and made shots, utilizing Silverman's rule \citep{silvermanDensityEstimationStatistics1986} for bandwidth estimation. The density estimation is conducted on a regularly spaced grid consisting of $201 \times 201$ points (see Figure \ref{fig:shoots_make_miss} for the estimated densities for Stephen Curry). We obtain observations of bivariate functional data (i.e., $P = 2$ functional features), where both features are defined on a two-dimensional rectangular domain, and observed on an equidistant grid of size $201 \times 201$.

We estimate the functional principal components using the \texttt{Gram}, \texttt{(Tensor) PCA} and \texttt{2D/1D B-Splines} methods. For the \texttt{Gram} method, we directly use the estimated densities to estimate the eigencomponents. For the \texttt{(Tensor) PCA} method, the estimation of the multivariate eigenimages is based on the univariate estimation of $30$ univariate eigenimages. The estimation of the (univariate) eigenimages is performed with the FCP-TPA where the smoothing parameters are chosen via cross-validation in $[10^{-5}, 10^5]$. For the \texttt{2D/1D B-splines} method, the two components are expanded in tensor products of $13 \times 13$ B-splines. We do not add a smoothing penalty in that case, as the smoothing step is performed in the kernel density estimation.

\begin{figure}
    \centering
    \includegraphics[width=\textwidth]{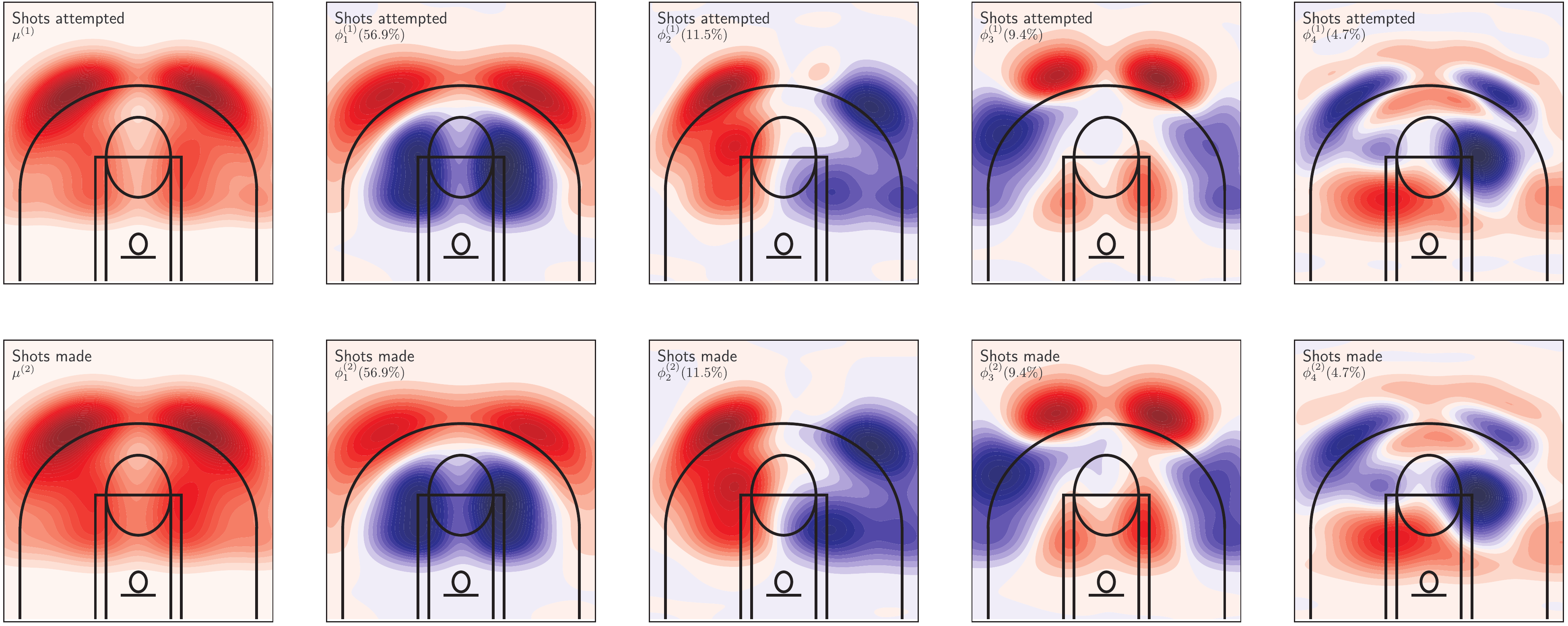}
    \caption{The estimated mean surfaces (first column) and the estimated eigenfunctions (second to fifth columns) for the shots dataset using the \texttt{Gram} method.}
    \label{fig:eigenfunctions_gram}
\end{figure}

Figure~\ref{fig:eigenfunctions_gram} shows the estimated mean surfaces and eigenfunctions for the \texttt{Gram} method. The functional principal components, as a representation of the deviation from the mean surface function, may take negative values, while densities can only take positive values. For presentation purposes, the functional principal components have been normalized between $-1$ and $1$. Blue areas correspond to negative values of the functional components, and thus contribute negatively to the density, while red area corresponds to positive values of the functional components, and thus contribute positively to the density. The decomposition of the density of made shots is similar that of attempted shots. The obtained functional components can be explained as different shooting behavior. The first component, accounting for $56.9\%$ of the variance explained, contrasts two-point and three-point shooting. A player with a positive score for $\phi_1$ will tend to take/attempt shots more behind the three-point line, while if he has a negative score, he will prefer shooting within the two-points zone. For the second component ($11.5\%$ of the variance explained), the contrast is between the left and right sides of the court. Similarly, if a player has a positive (resp. negative) score for $\phi_2$, he will prefer to shoot on the right (resp. left) side of the court while looking at the hoop. The third component ($9.4\%$ of the variance explained) contrasts shooting from the wing with shooting in the axis of the hoop. Positive (resp. negative) score players will shot in the axis of the hoop (resp. from the wing). The fourth component ($4.7\%$ of the variance explained) is more difficult to explain from a shooting behavior perspective. It may be linked to players having a preferred shooting location so that most of their shots are taken from the same position. The estimation of the eigencomponents with the \texttt{(Tensor) PCA} and \texttt{2D/1D B-Splines} methods are similar to the results with the \texttt{Gram} method and are thus provided in the Appendix \ref{sec:more_results}. 

\section{Discussion and conclusion} 
\label{sec:discussion}

MFPCA is a fundamental statistical tool for the analysis of multivariate functional data, which enables us to capture the variability in observations defined by multiple curves. In this paper, we have described the duality between rows and columns of a data matrix within the context of multivariate functional data. We have proposed to use this duality to estimate the eigencomponents of the covariance operator in multivariate functional datasets. By comparing the results of the three methods, we provide the researcher with guidelines for determining the most appropriate method within a range of functional data frameworks. 
Overall, our simulations showed that the \texttt{(Tensor) PCA}, \texttt{2D/1D B-Splines} and \texttt{Gram} methods give similar results in terms of the estimation of the eigenvalues, the eigenfunctions and reconstruction of the curves. Regarding the computation time, the use of the Gram matrix is faster in most cases. The only situation where the diagonalization of the covariance operator is quicker is when the number of observations is larger than the number of sampling points.
In conclusion, regarding the reconstruction error and computational complexity, if the data are defined on multi-dimensional domains (images) or the number of sampling points is much higher than the number of observations, we advise using the Gram matrix. Another advantage of the \texttt{Gram} method is that it does not require basis expansion, such as the \texttt{2D/1D B-splines} method, or estimate smoothing parameters, such as the \texttt{(Tensor) PCA} method when the data are observed without noise.

In practice, observations of (multivariate) functional data are often subject to noise. In this case, the diagonalization of the covariance operator seems preferable as the curves as to be smoothed beforehand to estimate the Gram matrix and yield similar results, except when the number of sampling is small where the \texttt{Gram} method is preferable. In the case of sparsely sampled functional data, we would advise to use the \texttt{Gram} method, using linear interpolation to estimate the inner-product matrix, as it does not require hyper-parameters estimation and gives slightly better results than the other methods.

Utilizing the Gram matrix enables the estimation of the number of components retained via the percentage of variance explained by the multivariate functional data, whereas the decomposition of the covariance operator necessitates the specification of the percentage of variance accounted for by each individual univariate feature. Specifying the percentage of variance explained for each feature does not guarantee recovery of the nominal percentage of variance explained for the multivariate functional data \citep{golovkineEstimationNumberComponents2023}. Although we have not investigated the extent to which this might be important, the duality relation derived in this work provides a direct solution to the problem. In settings where the univariate variance-explained cutoffs fail to retain the correct percentage of variance explained in multivariate functional data, the Gram matrix approach may be preferred.

The open-source implementation can be accessed at \url{https://github.com/StevenGolovkine/FDApy}, while scripts to reproduce the simulations and data analysis are at \url{https://github.com/FAST-ULxNUIG/geom_mfpca}.



\setcounter{equation}{0}
\renewcommand{\theequation}{A.\arabic{equation}}
\begin{appendix}

\section{Derivation of the equalities} 
\label{sec:derivation_of_the_inertia_of_the_clouds}

\subsection{Proof of Lemma \ref{lm:adjoint}} 
\label{sub:proof_of_lemma_lm:adjoint}

Using the definition of adjoint operators, we must prove that
\begin{equation}\label{eq:adjoint_op}
    \inRM{L_X(f)}{u} = \inHG{f}{L^\star_X(u)}, \quad\text{for all}\quad f \in \HH,\quad u \in \RR^N.
\end{equation}

\begin{proof}
For all $f \in \HH, u \in \RR^N$, we have that
\begin{align*}
    \inRM{L_X(f)}{u} &= L_X(f)^\top \mathbf{M} u \\
    &= \sum_{i = 1}^N\sum_{j = 1}^N \pi_i \sqrt{\pi_j} u_j \inH{X_i - \mu}{X_j - \mu}\inH{X_i - \mu}{f}, \\
    \inHG{f}{L^\star_X(u)} &= \inH{\Gamma f}{L^\star_X(u)} \\
    &= \sum_{p = 1}^P \int_{\TT{p}} (\Gamma f)^{(p)}(t_p) \left\{ \sum_{n = 1}^N \sqrt{\pi_n} u_n \left(\Xnp(t_p) - \mup{p}(t_p)\right) \right\} \dd t_p \\
    &= \sum_{p = 1}^P \int_{\TT{p}} \left\{\sum_{q = 1}^P \int_{\TT{q}} C_{p,q}(t_p, s_q)f^{(q)}(s_q) \dd s_q\right\} \left\{ \sum_{j = 1}^N \sqrt{\pi_j} u_j \left(X_j^{(p)}(t_p) - \mup{p}(t_p)\right) \right\} \dd t_p \\
    &= \sum_{i = 1}^N \sum_{j = 1}^N \pi_i \sqrt{\pi_j} u_j \left\{\sum_{p = 1}^P \int_{\TT{p}} \left(X_i^{(p)}(t_p) - \mup{p}(t_p)\right)\left(X_j^{(p)}(t_p) - \mup{p}(t_p)\right)\dd t_p\right\} \times \\
    &\qquad\qquad \left\{\sum_{q = 1}^P \int_{\TT{q}} \left(X_i^{(q)}(s_q) - \mup{q}(s_q)\right)f^{(q)}(s_q)\dd s_q\right\} \\
    &= \sum_{i = 1}^N\sum_{j = 1}^N \pi_i \sqrt{\pi_j} u_j \inH{X_i - \mu}{X_j - \mu}\inH{X_i - \mu}{f}.
\end{align*}
So, the equality \eqref{eq:adjoint_op} is proved and we conclude that $L^\star_X$ is the adjoint operator of $L_X$.    
\end{proof}


\subsection{Derivation of the inertia of the clouds} 
\label{sub:derivation_of_the_inertia_of_the_clouds}

Recall that 
\begin{equation}\label{eq:var_appendix}
    \Var\{\Xp{p}(t_p)\} = \sum_{n = 1}^N \pi_n \{\Xnp(t_p)\}^2 - \{\mup{p}(t_p)\}^2 \quad\text{where}\quad \mup{p}(t_p) = \sum_{n = 1}^N \pi_n\Xnp(t_p), \quad t_p \in \TT{p}.
\end{equation}
We first show that the total inertia of the cloud of individuals $\CP$ using the distance $d$ is given by equalities in~\eqref{eq:inertia}:
$$\sum_{n = 1}^N \pi_n d^2(\pobs{M}_n, \Gmu) = \frac{1}{2}\sum_{i = 1}^N \sum_{j = 1}^N \pi_i \pi_j d^2(\pobs{M}_i, \pobs{M}_j) = \sum_{p = 1}^P \int_{\TT{p}}\Var{\Xp{p}(t_p)} \dd t_p.$$

\begin{proof}
    \begin{align*}
    \sum_{n = 1}^N \pi_n d^2(\pobs{M}_n, \Gmu) 
    &= \sum_{n = 1}^N \pi_n \sum_{p = 1}^P\normLp{\Xnp - \mu^{(p)}}^2\\
    &= \sum_{p = 1}^P\left(\sum_{n = 1}^N \pi_n \normLp{\Xnp}^2 - \normLp{\mup{p}}^2\right) \\
    &= \sum_{p = 1}^P \int_{\TT{p}}\Var{\Xp{p}(t_p)} \dd t_p, \\
\sum_{i = 1}^N \sum_{j = 1}^N \pi_i \pi_j d^2(\pobs{M}_i, \pobs{M}_j) &= \sum_{i = 1}^N \sum_{j = 1}^N \pi_i \pi_j \sum_{p = 1}^P \normLp{\Xp{p}_i - \Xp{p}_j}^2\\
    &= \sum_{p = 1}^P \left(2\sum_{i = 1}^N \pi_i\normLp{\Xp{p}_i}^2 - 2\sum_{i = 1}^N \sum_{j = 1}^N \pi_i\pi_j\inLp{\Xp{p}_i}{\Xp{p}_j}\right) \\
    &= 2\sum_{p = 1}^P \int_{\TT{p}}\Var{\Xp{p}(t_p)} \dd t_p.
\end{align*}
The equalities in Equation~\eqref{eq:inertia} are shown.
\end{proof}
Next, we derive the total inertia of the cloud of individuals $\CP$ using the distance $d_\Gamma$, given by equalities in~\eqref{eq:inertia_CP}:
$$\sum_{n = 1}^N \pi_n d_\Gamma^2(\pobs{M}_n, \Gmu) = \frac{1}{2}\sum_{i = 1}^N \sum_{j = 1}^N \pi_i \pi_j d_\Gamma^2(\pobs{M}_i, \pobs{M}_j) = \sum_{p = 1}^P \int_{\TT{p}} \normH{C_{p \cdot}(t_p, \cdot)}^2 \dd t_p. $$

\begin{proof}
\begin{align}
   \sum_{n = 1}^N \pi_n d_\Gamma^2(\pobs{M}_n, \Gmu) &= \sum_{n = 1}^N \pi_n \inHG{X_n - \mu}{X_n - \mu} \\
   &= \sum_{n = 1}^N \pi_n \normHG{X_n}^2 - \normHG{\mu}^2 \\
   &= \sum_{p = 1}^P \left(\sum_{n = 1}^N \pi_n \normLp{\Xnp}_{\Gamma}^2 - \normLp{\mup{p}}_{\Gamma}^2\right) \\
   &= \sum_{p = 1}^P \sum_{q = 1}^P \int_{\TT{p}} \int_{\TT{q}} C_{p,q}(t_p, s_q)C_{p,q}(t_p, s_q) \dd s_q \dd t_p \\
   &= \sum_{p = 1}^P \int_{\TT{p}} \normH{C_{p \cdot}(t_p, \cdot)}^2 \dd t_p, \\
   \frac{1}{2}\sum_{i = 1}^N \sum_{j = 1}^N \pi_i \pi_j d_\Gamma^2(\pobs{M}_i, \pobs{M}_j)&= \sum_{i = 1}^N \pi_i \normHG{X_i}^2 - \sum_{i = 1}^N \sum_{j = 1}^N \pi_i\pi_j \inHG{X_i}{X_j} \\
    &= \sum_{i = 1}^N \pi_i \normHG{X_i}^2 - \normHG{\mu}^2 \\
    &= \sum_{p = 1}^P \int_{\TT{p}} \normH{C_{p \cdot}(t_p, \cdot)}^2 \dd t_p. 
\end{align}
The equalities in Equation~\eqref{eq:inertia_CP} are shown.
\end{proof}
Finally, we derive the inertia of the cloud of features $\CN$ using the distance $\mathsf{d}$, given by equalities in~\eqref{eq:inertia_CN}:
$$\sum_{n = 1}^N \pi_n \mathsf{d}^2(\pfea{M}_n, \Gfea) = \frac{1}{2}\sum_{i = 1}^N \sum_{j = 1}^N \pi_i \pi_j \mathsf{d}^2(\pfea{M}_i, \pfea{M}_j) = \sum_{p = 1}^P \int_{\TT{p}} \normH{C_{p \cdot}(t_p, \cdot)}^2 \dd t_p.$$

\begin{proof}
\begin{align}
\sum_{n = 1}^N \pi_n \mathsf{d}^2(\pfea{M}_n, \Gfea) &= \sum_{i = 1}^N \sum_{j = 1}^N \pi_i \pi_j \inH{X_i - \mu}{X_j - \mu}\inH{X_i - \mu}{X_j - \mu} \\
&= \sum_{p = 1}^P \sum_{q = 1}^P \int_{\TT{p}} \int_{\TT{q}} C_{p,q}(t_p, s_q)C_{p,q}(t_p, s_q) \dd s_q \dd t_p \\
&= \sum_{p = 1}^P \int_{\TT{p}} \normH{C_{p \cdot}(t_p, \cdot)}^2 \dd t_p,\\
\sum_{i = 1}^N \sum_{j = 1}^N \pi_i \pi_j \mathsf{d}^2(\pfea{M}_i, \pfea{M}_j) &= \sum_{i = 1}^N \sum_{j = 1}^N \sum_{n = 1}^N \pi_i \pi_j \pi_n \inH{X_n - \mu}{X_i - X_j}\inH{X_n - \mu}{X_i - X_j} \\
&= 2\sum_{i = 1}^N \sum_{j = 1}^N \pi_i \pi_j \inH{X_i - \mu}{X_j - \mu}\inH{X_i - \mu}{X_j - \mu} \\
&= 2\sum_{p = 1}^P \int_{\TT{p}} \normH{C_{p \cdot}(t_p, \cdot)}^2 \dd t_p.
\end{align}
The equalities in Equation~\eqref{eq:inertia_CN} are shown.    
\end{proof}



\section{Derivation of the eigencomponents} 
\label{sec:derivation_of_the_eigencomponents}

\subsection{General framework} 
\label{sub:general_framework}
In this section, we calculate the relationships between the eigenelements of the covariance operator $\Gamma$ and the ones of the Gram matrix $\mathbf{M}$ of a functional dataset. We then prove the equalities~\eqref{eq:eigenvalues_relation_p},~\eqref{eq:eigenfunction_relation_p} and~\eqref{eq:scores_relation_p}.

Using the Hilbert-Schmidt theorem, there exists a complete orthonormal basis of eigenvectors $\{\boldsymbol{u}_k\}_{1 \leq k \leq N}$ of the inner-product matrix $\mathbf{M}$ such that
\begin{equation}\label{eq:eigen_inner_prod_p}
    \mathbf{M}\boldsymbol{u}_k = l_k\boldsymbol{u}_k.
\end{equation}
Let $X = \left(X_1 - \mu, \dots, X_N - \mu\right)^\top$ and denote $\widetilde{X} = \text{diag}\{\sqrt{\pi_1}, \dots, \sqrt{\pi_N}\}X$, the matrix of weighted observations. Recall that, in the case of $P$-dimensional process, the realisations of the process $X_n,~n = 1, \cdots, N$ and $\mu$ are vectors of functions of length $P$, and thus $X$ (and $\widetilde{X}$) is a matrix of functions of size $N \times P$. By left multiplying Equation~\eqref{eq:eigen_inner_prod_p} by $\widetilde{X}^\top$, we obtain
\begin{equation}\label{eq:eigen_inner_prod_left}
    \widetilde{X}^\top \mathbf{M} \boldsymbol{u}_k = l_k \widetilde{X}^\top \boldsymbol{u}_k.
\end{equation} 
Expanding Equation~\eqref{eq:eigen_inner_prod_left}, for each component $p = 1, \dots, P$, we have,
\begin{equation}\label{eq:inner_prod_p}
    \sum_{i = 1}^N \sum_{j = 1}^N \pi_i \sqrt{\pi_j}[\boldsymbol{u}_{k}]_j\mkern-4mu\left\{X_i^{(p)}(\cdot) - \mu^{(p)}(\cdot)\right\}\inH{X_i - \mu}{X_j - \mu} = l_k \mkern-5mu\sum_{n = 1}^N \mkern-4mu\sqrt{\pi_n}[\boldsymbol{u}_{k}]_n\mkern-4mu\left\{\Xnp(\cdot) - \mu^{(p)}(\cdot)\right\}.
\end{equation}
Here and in the following, we note $[a]_p$ the $p$th entry of the vector $a$. Starting from the left side of Equation~\eqref{eq:inner_prod_p}, we get
\begin{align}\label{eq:inner_prod_p_left}
[\widetilde{X}^\top \mathbf{M} \boldsymbol{u}_k]_p &= \sum_{i = 1}^N \sum_{j = 1}^N \pi_i \sqrt{\pi_j} [\boldsymbol{u}_{k}]_j \left\{X_i^{(p)}(\cdot) - \mu^{(p)}(\cdot)\right\}\inH{X_i - \mu}{X_j - \mu}\\
&= \sum_{q = 1}^P \int_{\TT{q}} \sum_{i = 1}^N \pi_i\left\{X_i^{(p)}(\cdot) - \mu^{(p)}(\cdot)\right\} \left\{X_i^{(q)}(s_q) - \mu^{(q)}(s_q)\right\}  \\
&\qquad\qquad \times \sum_{j = 1}^N \sqrt{\pi_j}[\boldsymbol{u}_{k}]_j \left\{X_j^{(q)}(s_q) - \mu^{(q)}(s_q)\right\} \dd s_q \\
&= \sum_{q = 1}^P \int_{\TT{q}} C_{p,q}(\cdot, s_q)\sum_{j = 1}^N \sqrt{\pi_j}[\boldsymbol{u}_{k}]_j \left\{X_j^{(q)}(s_q) - \mu^{(q)}(s_q)\right\} \dd s_q \\
&= \sum_{j = 1}^N \inH{C_{p \cdot}(\cdot, \cdot)}{\sqrt{\pi_j}[\boldsymbol{u}_{k}]_j \left\{X_j - \mu\right\}} \\
&= \Gamma\left(\sum_{j = 1}^N \sqrt{\pi_j}[\boldsymbol{u}_{k}]_j \left\{X_j - \mu\right\} \right)^{\mkern-9mu(p)}\mkern-18mu(\cdot),
\end{align}
and, starting from the right side of Equation~\eqref{eq:inner_prod_p},
\begin{equation}\label{eq:inner_prod_p_right}
    [l_k \widetilde{X}^\top \boldsymbol{u}_k]_p = l_k \sum_{n = 1}^N \sqrt{\pi_n}[\boldsymbol{u}_{k}]_n \left\{\Xnp(\cdot) - \mu^{(p)}(\cdot)\right\}.
\end{equation}
From Equation~\eqref{eq:inner_prod_p_left} and Equation~\eqref{eq:inner_prod_p_right}, we obtain
\begin{equation}
    \Gamma\left(\sum_{j = 1}^N \sqrt{\pi_j}[\boldsymbol{u}_{k}]_j \left\{X_j - \mu\right\}\right)^{\mkern-9mu(p)}\mkern-18mu(\cdot) = l_k \sum_{n = 1}^N \sqrt{\pi_n}[\boldsymbol{u}_{k}]_n \left\{\Xnp(\cdot) - \mu^{(p)}(\cdot)\right\}, \quad\text{for all}~ p = 1, \dots, P.
\end{equation}
By identification in Equation~\eqref{eq:eigendecomposition}, we find that, for each components $p$,
\begin{equation}\label{eq:eigen_estimation}
\lambda_k = l_k \quad\text{and}\quad \phi_k^{(p)}(\cdot) = \sum_{n = 1}^N \sqrt{\pi_n}[\boldsymbol{u}_{k}]_n \left\{\Xnp(\cdot) - \mu^{(p)}(\cdot)\right\}, \quad k \geq 1.
\end{equation}
For $k \geq 1$, the norm of the eigenfunction is computed as the following:
\begin{align*}
\normH{\phi_k}^2 &= \sum_{i = 1}^N \sum_{j = 1}^N \sqrt{\pi_i\pi_j}[\boldsymbol{u}_{k}]_i [\boldsymbol{u}_{k}]_j\inH{X_i - \mu}{X_j - \mu} = \sum_{i = 1}^N [\boldsymbol{u}_{k}]_i \sum_{j = 1}^N \mathbf{M}_{ij} [\boldsymbol{u}_{k}]_j \\
    &= \sum_{i = 1}^N [\boldsymbol{u}_{k}]_i l_k [\boldsymbol{u}_{k}]_i = l_k \normR{\boldsymbol{u}_k}^2 = l_k. \\
\end{align*}
Therefore, in order to have an orthonormal basis of eigenfunctions, we normalise the eigenfunctions $\phi_k$ from Equation~\eqref{eq:eigen_estimation} by $1 / \sqrt{l_k}$.
Concerning the estimation of the scores, for $n = 1, \dots, N$, for $k \geq 1$, we have
\begin{align}
    \mathfrak{c}_{nk} &= \inH{X_n - \mu}{\phi_k} = \frac{1}{\sqrt{l_k}}\sum_{j = 1}^N \sqrt{\pi_j}[\boldsymbol{u}_{k}]_j \inH{X_n - \mu}{X_j - \mu}\\
    &= \frac{1}{\sqrt{l_k\pi_n}}\sum_{j = 1}^N [\boldsymbol{u}_{k}]_j \mathbf{M}_{nj} = \sqrt{\frac{l_k}{\pi_n}}[\boldsymbol{u}_{k}]_n.\\
\end{align}

If we assume that the observations are equally weighted, i.e., $\pi_n = 1 / N, n = 1, \dots, N$, we get the equalities~\eqref{eq:eigenvalues_relation_p},~\eqref{eq:eigenfunction_relation_p} and~\eqref{eq:scores_relation_p}.



\section{More results} 
\label{sec:more_results}

\subsection{Simulation} 
\label{sub:simulation}

\begin{figure}
    \centering
    \includegraphics[width=0.95\textwidth]{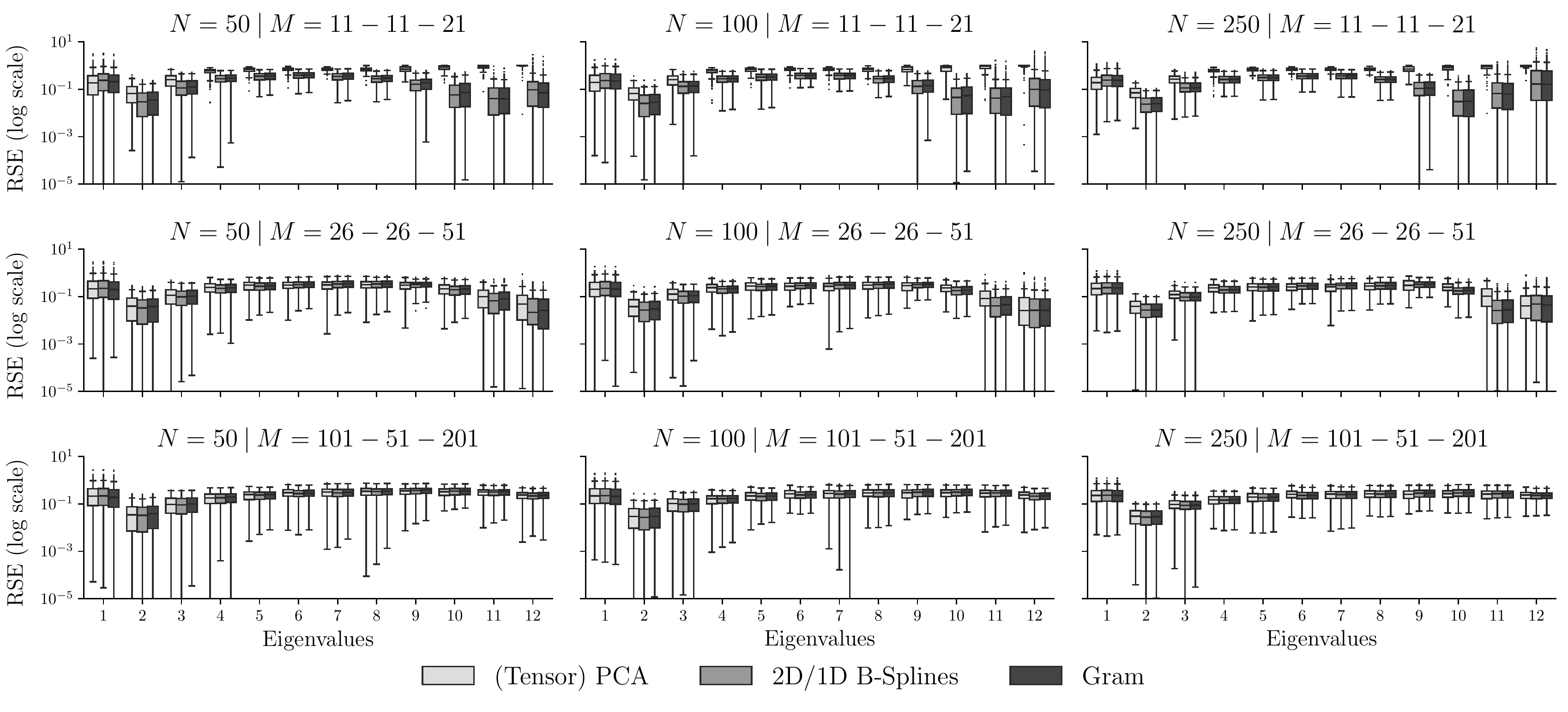}
    \caption{RSE for the estimated eigenvalues for each method in the noisy case. $N$ is the number of observations, $M$ is the number of sampling points per curve (the first two numbers are for the images and the last one is for the curves).}
    \label{fig:logAE_mfd_1d_noise}
\end{figure}

\begin{figure}
     \centering
    \includegraphics[width=0.95\textwidth]{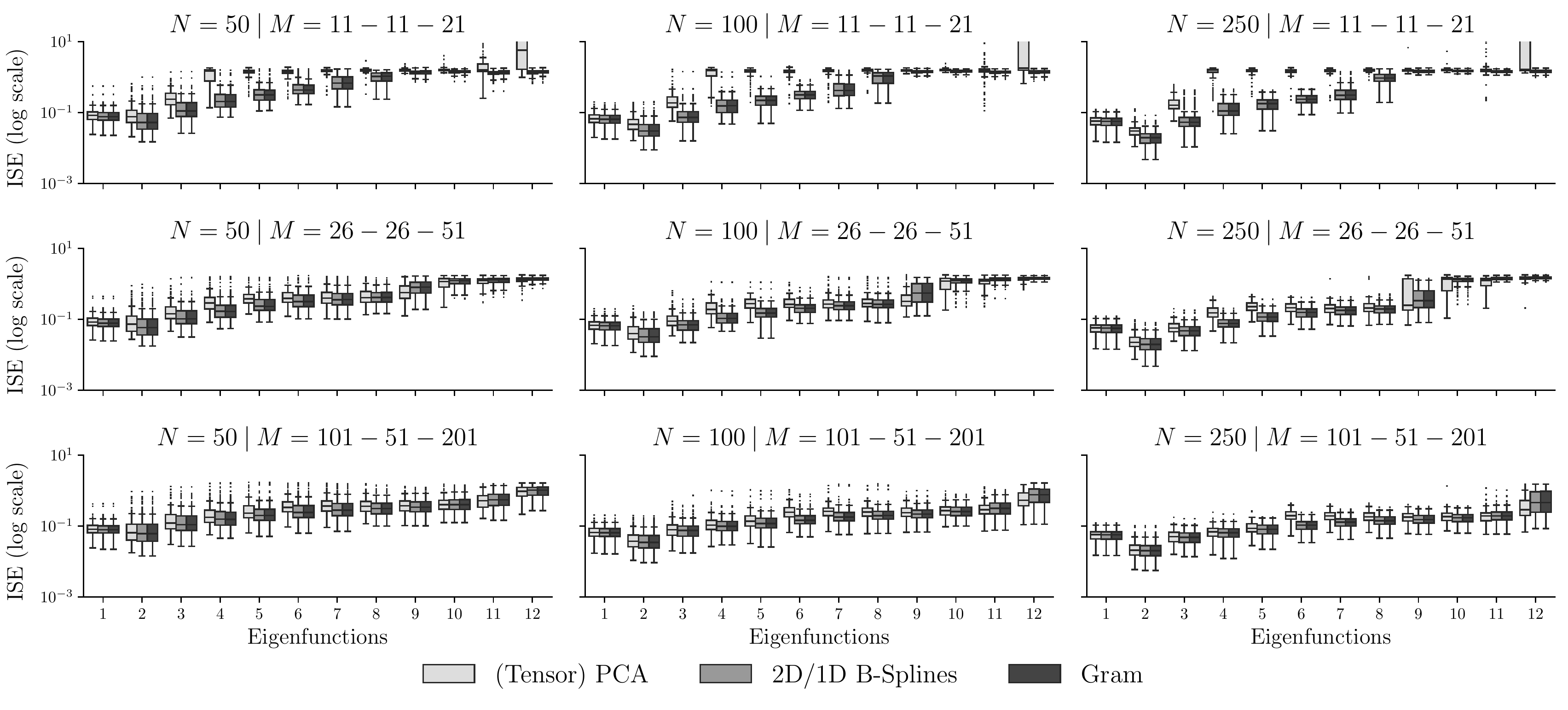}
    \caption{ISE for the estimated eigenfunctions for each method in the noisy case. $N$ is the number of observations, $M$ is the number of sampling points per curve (the first two numbers are for the images and the last one is for the curves).}
    \label{fig:ise_mfd_1d_noise}
\end{figure}

\begin{figure}
     \centering
     \includegraphics[width=0.95\textwidth]{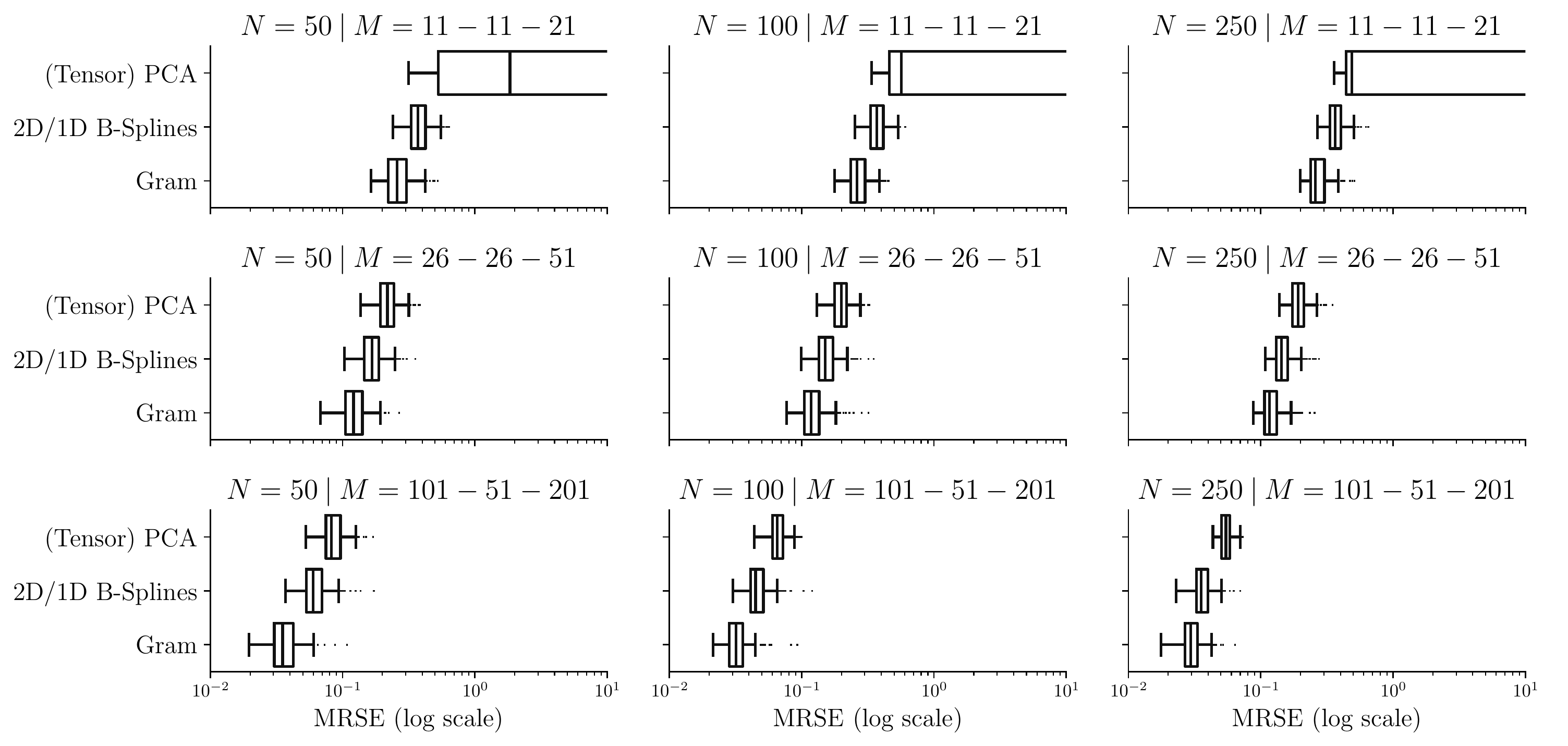}
    \caption{MRSE for the reconstructed curves for each method in the noisy case. $N$ is the number of observations, $M$ is the number of sampling points per curve (the first two numbers are for the images and the last one is for the curves).}
    \label{fig:mise_mfd_1d_noise}
\end{figure}

We present the results of the simulations in the sparse and noisy cases. Figures \ref{fig:logAE_mfd_1d_noise}, \ref{fig:ise_mfd_1d_noise} and \ref{fig:mise_mfd_1d_noise} present the boxplots of the RSE, ISE and MRSE, respectively, for the noisy case. To generate noisy data, we consider the model \eqref{eq:model_error} where $\sigma^2 = 0.25$. For the \texttt{(Tensor) PCA} method, we first smooth the 2-dimensional data using P-Splines smoothing and estimate a smooth version of the mean and covariance functions for the one-dimensional data using P-Splines smoothing. For the \texttt{2D/1D B-Splines} and \texttt{Gram} methods, all the observations have been smoothed using P-Splines smoothing beforehand. In every case, the penalty involved in P-Splines smoothing has been estimated using cross-validation.

\begin{figure}
    \centering
    \includegraphics[width=0.8\textwidth]{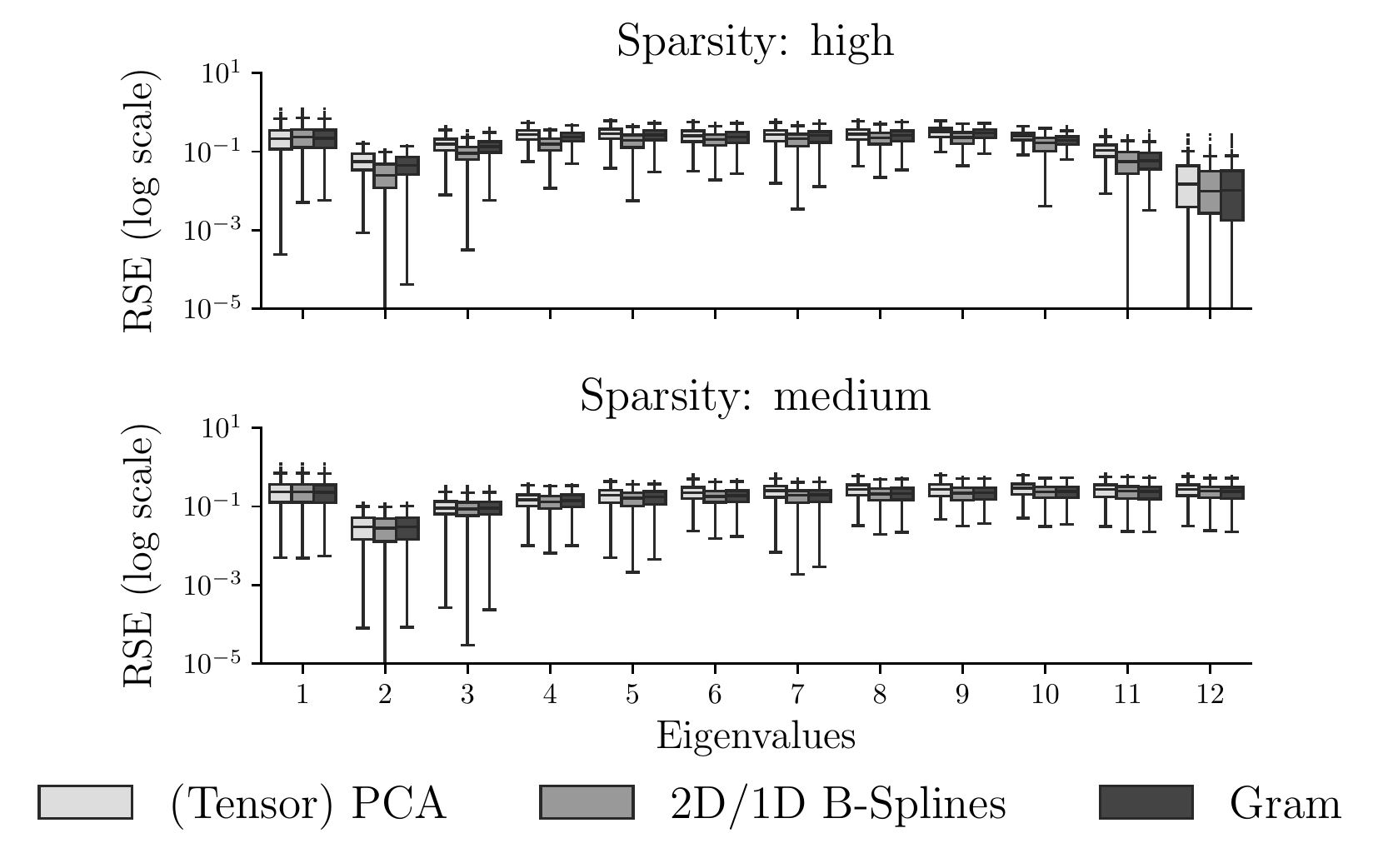}
    \caption{RSE for the estimated eigenvalues for each method in the sparse case. We set $N = 250$, $M^{(1)} = 101 \times 51$ and $M^{(2)} = 201$. For the case of medium sparsity, we remove $50\%-70\%$ of the sampling points and for the case of high sparsity, we remove $90\%-95\%$ of the sampling points.}
    \label{fig:logAE_mfd_1d_sparse}
\end{figure}

\begin{figure}
     \centering
    \includegraphics[width=0.95\textwidth]{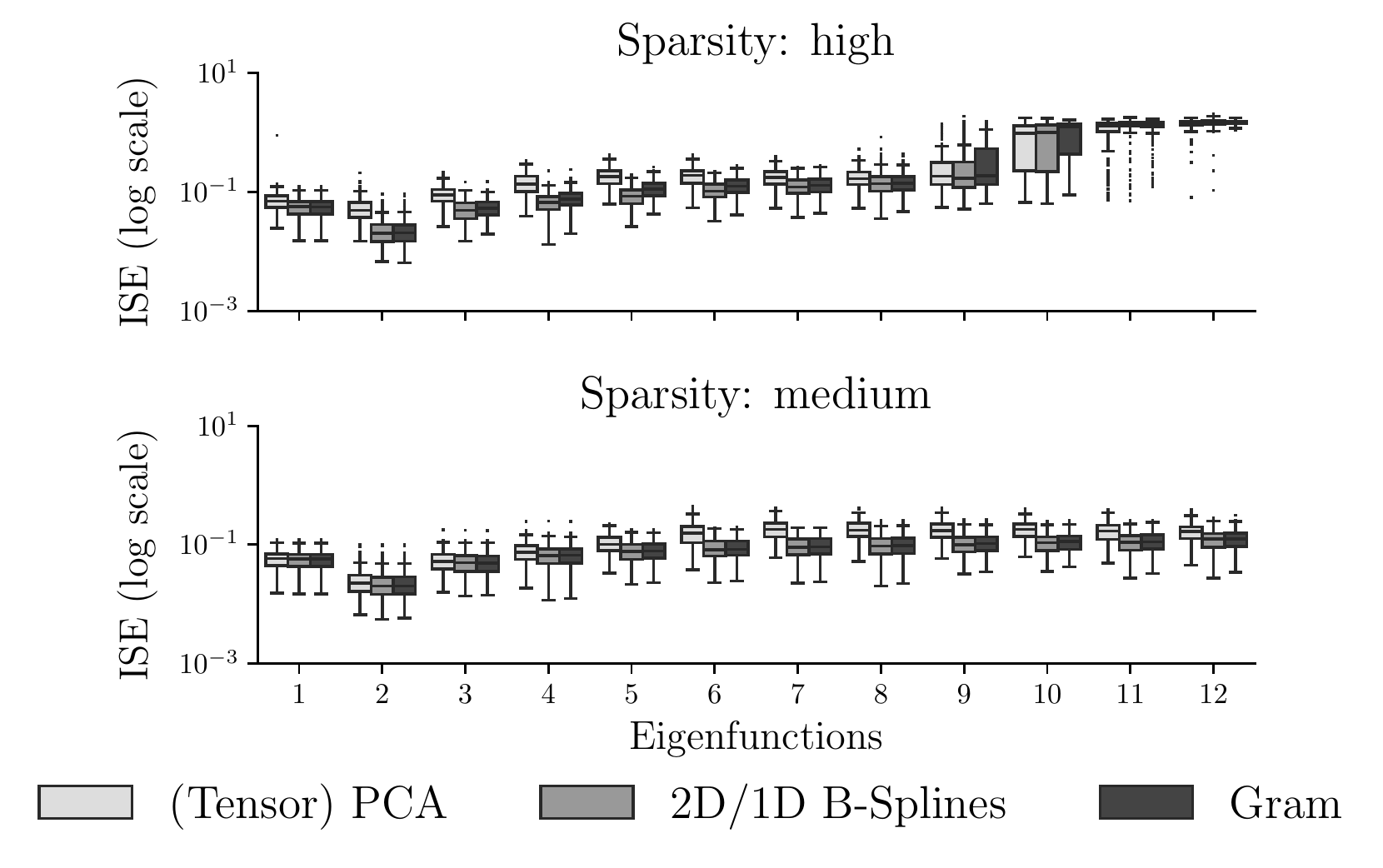}
    \caption{ISE for the estimated eigenfunctions for each method in the sparse case. We set $N = 250$, $M^{(1)} = 101 \times 51$ and $M^{(2)} = 201$. For the case of medium sparsity, we remove $50\%-70\%$ of the sampling points and for the case of high sparsity, we remove $90\%-95\%$ of the sampling points.}
    \label{fig:ise_mfd_1d_sparse}
\end{figure}

\begin{figure}
     \centering
     \includegraphics[width=0.95\textwidth]{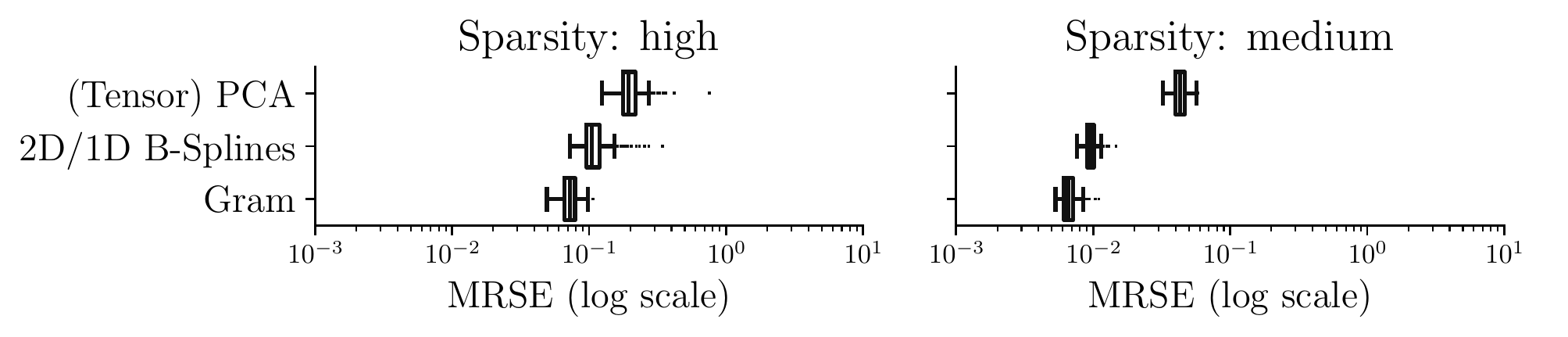}
    \caption{MRSE for the reconstructed curves for each method in the sparse case. We set $N = 250$, $M^{(1)} = 101 \times 51$ and $M^{(2)} = 201$. For the case of medium sparsity, we remove $50\%-70\%$ of the sampling points and for the case of high sparsity, we remove $90\%-95\%$ of the sampling points.}
    \label{fig:mise_mfd_1d_sparse}
\end{figure}

Figures \ref{fig:logAE_mfd_1d_sparse}, \ref{fig:ise_mfd_1d_sparse} and \ref{fig:mise_mfd_1d_sparse} present the boxplots of the RSE, ISE and MRSE, respectively, for the sparse case. In this case, we only consider $N = 250$, $M^{(1)} = 101 \times 51$, $M^{(2)} = 201$ and no noise. We consider medium sparsity, where $50\%-70\%$ of the sampling points have been removed and high sparsity, where $90\%-95\%$ of the sampling points have been removed. For the \texttt{(Tensor) PCA} method, we first interpolate the 2-dimensional data to have the observations reguarlarly sampled on a grid and estimate a smooth version of the mean and covariance functions for the one-dimensional data using P-Splines smoothing. For the \texttt{2D/1D B-Splines} method, all the observations have been smoothed using P-splines smoothing with a penalty estimated by cross-validation. For the \texttt{Gram} method, the observations have been linearly interpolated to estimate the inner-product matrix.

\subsection{Application} 
\label{sub:application}

\begin{figure}
    \centering
    \includegraphics[width=\textwidth]{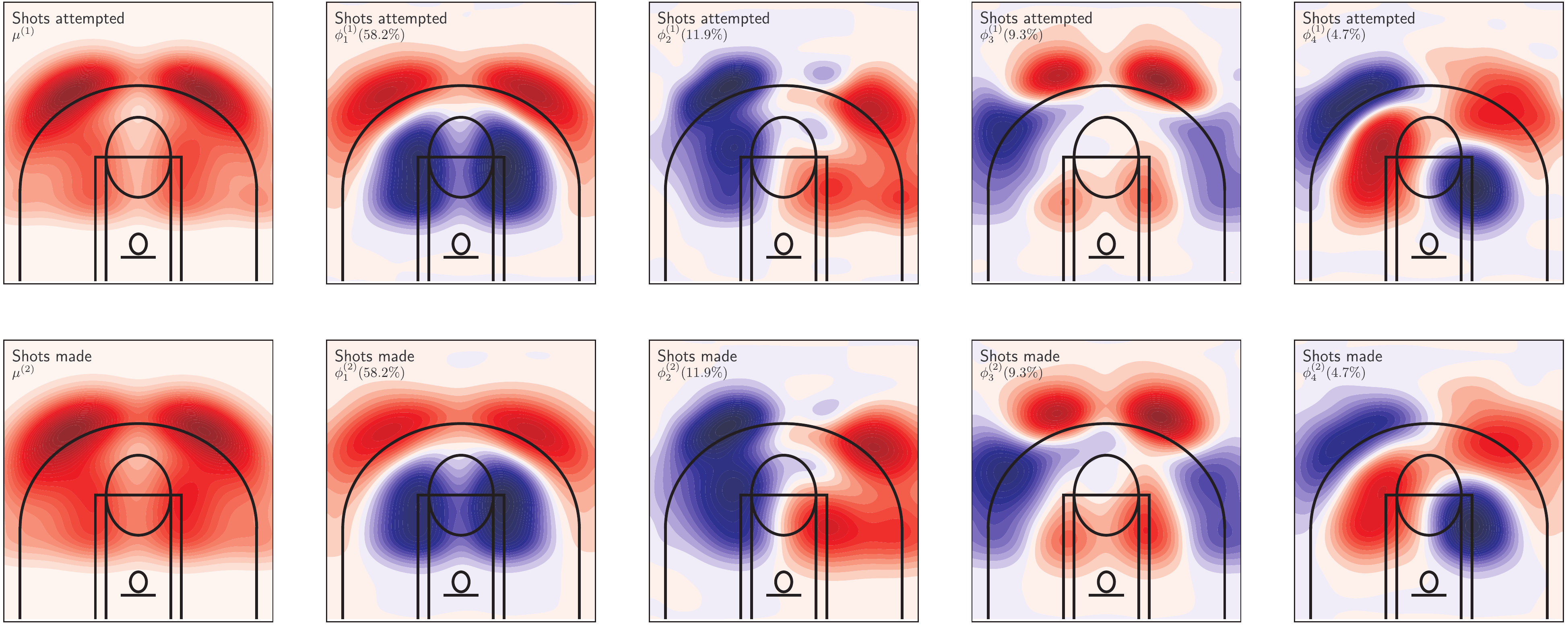}
    \caption{The estimated mean surfaces (first column) and the estimated eigenfunctions (second to fifth columns) for the shots dataset using the \texttt{(Tensor) PCA} method.}
    \label{fig:eigenfunctions_fcptpa}
\end{figure}

\begin{figure}
    \centering
    \includegraphics[width=\textwidth]{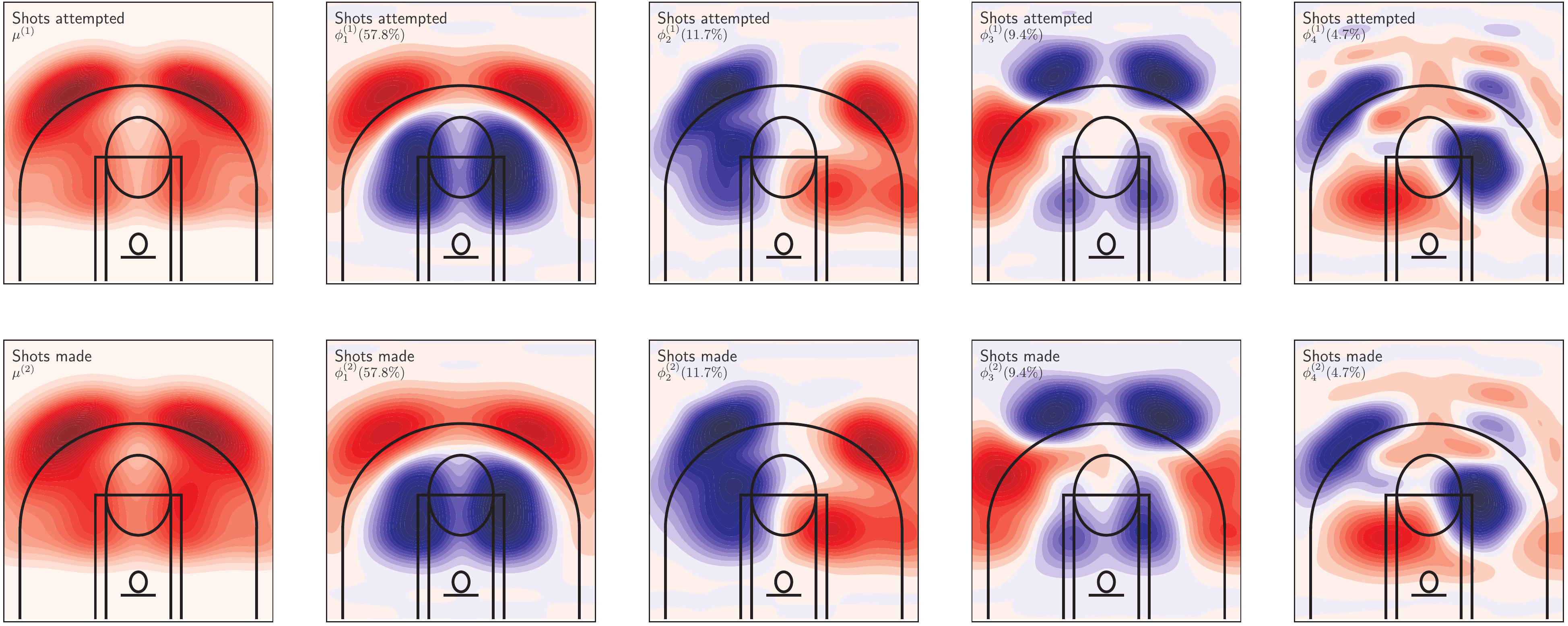}
    \caption{The estimated mean surfaces (first column) and the estimated eigenfunctions (second to fifth columns) for the shots dataset using the \texttt{2D/1D B-splines} method.}
    \label{fig:eigenfunctions_psplines}
\end{figure}

Here, we present the estimation of the eigencomponents of the shooting data using the \texttt{(Tensor) PCA} (see Figure~\ref{fig:eigenfunctions_fcptpa}) and \texttt{2D/1D B-Splines} (see Figure~\ref{fig:eigenfunctions_psplines}) methods. The results are similar to those found using the \texttt{Gram} method for the eigenfunctions and the percentage of variance explained for each eigenfunctions.

\end{appendix}

\section*{Acknowledgment}

S. Golovkine, A. J. Simpkin and N. Bargary are partially supported by Science Foundation Ireland under Grant No. 19/FFP/7002 and co-funded under the European Regional Development Fund. E. Gunning is supported in part Science Foundation Ireland (Grant No. 18/CRT/6049) and co-funded under the European Regional Development Fund.

\bibliographystyle{apalike}
\bibliography{biblio}

\makeatletter\@input{supplement_ref.tex}\makeatother

\end{document}


\maketitle

In this Supplementary Material, we provide insights for when the data are already decomposed in a basis, e.g. Fourier or polynomials. In particular, we explain how to perform MFPCA as described in Section~\ref{sec:functional_principal_components_analysis} in the main text.

\renewcommand{\theequation}{SM.\arabic{equation}}

\section{Basis decomposition} 
\label{sub:basis_decomposition}

In many practical situations, functional data are noisy and only observed at specific time points. To extract the underlying functional features of the data, smoothing and interpolation techniques are commonly employed. These techniques involve approximating the true underlying function generating the data by a finite-dimensional set of basis functions. Assume that for each feature $p = 1, \dots, P$, there exists a set of basis of functions $\Psi^{(p)} = \{\psi_k^{(p)}\}_{1 \leq k \leq K_p}$ such that each feature of each curve $n = 1, \dots, N$ can be expanded using the basis:
\begin{equation}\label{eq:curve_basis_expansion}
\Xnp(t_p) = \sum_{k = 1}^{K_p} c^{(p)}_{nk}\psi_k^{(p)}(t_p), \quad t_p \in \TT{p},
\end{equation}
where $\{c^{(p)}_{nk}\}_{1 \leq k \leq K_p}$ is a set of coefficients for feature $p$ of observation $n$. We denote by $\overline{c}_k^{(p)} = \sum_{n = 1}^N \pi_n c^{(p)}_{nk}$ the mean coefficient of feature $p$ corresponding to the $k$th basis function.
The $p$th feature of the mean function can be then expanded in the same basis as:
\begin{equation}
    \hatmup{p}(t_p) = \sum_{k = 1}^{K_p} \overline{c}_k^{(p)}\psi_k^{(p)}(t_p), \quad t_p \in \TT{p}.
\end{equation}
Similarly, the covariance function of the $p$th and $q$th features is given by:
\begin{equation}
    \widehat{C}_{p,q}(s_p, t_q) = \sum_{k = 1}^{K_p} \sum_{l = 1}^{K_q} \left(\sum_{n = 1}^N \pi_n c^{(p)}_{nk}c^{(q)}_{nl} - \overline{c}_k^{(p)}\overline{c}_l^{(q)}\right)\psi_k^{(p)}(s_p)\psi_l^{(q)}(t_q), \quad s_p \in \TT{p},\quad t_q \in \TT{q}.
\end{equation}
These formulas can be written in matrix form as follows. For $\pointt \in \TT{}$, we have that $X(\pointt) = \mathbf{C}\Psi(\pointt)$ where $X(\pointt)$ is a $N \times P$ matrix with entries $\Xnp(t_p),~t_p \in \TT{p},~1 \leq p \leq P,~1 \leq n \leq N$,
\begin{equation}
    \mathbf{C} = \begin{pmatrix}
            \mathbf{C}^{(1)} & \cdots & \mathbf{C}^{(P)} \\
        \end{pmatrix}, \quad \text{and}\quad
    \Psi(\pointt) = \text{diag}\{\Psi^{(1)}(t_1), \dots, \Psi^{(P)}(t_P)\},
\end{equation}
where
\begin{equation}
\mathbf{C}^{(p)} = \begin{pmatrix}
    c^{(p)}_{11} & \cdots & c^{(p)}_{1K_p} \\
    \vdots & \ddots & \vdots \\
    c^{(p)}_{N1} & \cdots & c^{(p)}_{NK_p}
\end{pmatrix} \\
\quad \text{and}\quad
\Psi^{(p)}(t_p) = \begin{pmatrix}
    \psi_1^{(p)}(t_p) \\
    \vdots \\
    \psi_{K_p}^{(p)}(t_p)
\end{pmatrix}.
\end{equation}
Using the basis expansion and denoting $\Pi^\top = (\pi_1, \dots, \pi_N)$, the mean and covariance functions are given by
\begin{equation}
    \widehat{\mu}(\pointt) = \Psi(\pointt)^\top \mathbf{C}^\top\Pi \quad\text{and}\quad \widehat{C}(\points, \pointt) = \Psi(\points)^\top \mathbf{C}^\top \left(\text{diag}\{
        \pi_1, \dots, \pi_N\} - \Pi\Pi^\top\right)\mathbf{C} \Psi(\pointt).
\end{equation}
Finally, we denote by $\mathbf{W}$ the matrix of inner products of the functions in the basis $\Psi$. The matrix $\mathbf{W}$ is a block-diagonal matrix such that $\mathbf{W} = \text{blockdiag}\{\mathbf{W}^{(1)}, \dots, \mathbf{W}^{(P)}\}$ where each entry is given by
\begin{equation}
    \mathbf{W}_{k, l}^{(p)} = \inLp{\psi_k^{(p)}}{\psi_l^{(p)}}, \quad 1 \leq k, l \leq K_p, \quad 1 \leq p \leq P.
\end{equation}
We remark that, if the basis $\Psi$ is an orthonormal basis, the matrix $\mathbf{W}$ is equal to the identity matrix of size $\sum_{p = 1}^P K_p$.
Using the expansion of the data into the basis of functions $\Psi$, the inner-product matrix $\mathbf{M}$ is written 
\begin{equation}\label{eq:gram_matrix_basis}
    \mathbf{M} = \text{diag}\{
        \sqrt{\pi_1}, \dots, \sqrt{\pi_N}\}\left(\mathrm{I}_{\!N} - \mathbf{1}_{\!N}\Pi^\top\right) \mathbf{C} \mathbf{W} \mathbf{C}^\top \left(\mathrm{I}_{\!N} - \Pi\mathbf{1}_{\!N}^\top\right)\text{diag}\{
        \sqrt{\pi_1}, \dots, \sqrt{\pi_N}\}
\end{equation}
where $\mathrm{I}_{\!N}$ is the identity matrix of size $N$ and $\mathbf{1}_{\!N}$ is a vector of $1$ of length $N$.


\section{MFPCA with a basis expansion} 
\label{sub:with_a_basis_expansion}

In this section, we assume that the observations are expanded into a basis of functions, as explained in Section~\ref{sub:basis_decomposition}. Using the expansion of the data into the basis of function $\Psi$ and $\mathbf{W}$, the matrix of inner products of the functions in the basis $\Psi$, we write \eqref{eq:gram_matrix_basis} as
\begin{equation}
    \mathbf{M} = \left(\text{diag}\{
        \sqrt{\pi_1}, \dots, \sqrt{\pi_N}\}\left(\mathrm{I}_{\!N} - \mathbf{1}_{\!N}\Pi^\top\right) \mathbf{C}\mathbf{W}^{1/2}\right)\left(\text{diag}\{
        \sqrt{\pi_1}, \dots, \sqrt{\pi_N}\}\left(\mathrm{I}_{\!N} - \mathbf{1}_{\!N}\Pi^\top\right) \mathbf{C}\mathbf{W}^{1/2}\right)^\top.
\end{equation}
We note
\begin{equation}
    \mathbf{A} = \text{diag}\{\sqrt{\pi_1}, \dots, \sqrt{\pi_N}\}\left(\mathrm{I}_{\!N} - \mathbf{1}_{\!N}\Pi^\top\right) \mathbf{C}\mathbf{W}^{1/2},
\end{equation}
such that $\mathbf{M} = \mathbf{A}\mathbf{A}^\top$.
We also assume that $\phi_1, \phi_2, \dots$ the eigenfunctions of the covariance operator $\Gamma$ have a decomposition into the basis $\Psi$
\begin{equation}
    \phi_k(\cdot) = 
        \begin{pmatrix} 
            \phi_k^{(1)}(\cdot) \\
            \vdots \\
            \phi_k^{(P)}(\cdot)
        \end{pmatrix} = 
        \begin{pmatrix} 
            \psi^{(1) \top}(\cdot) b_{1k} \\
            \vdots \\
            \psi^{(P) \top}(\cdot) b_{Pk}
        \end{pmatrix}, \quad\text{where}\quad
        b_{pk} = \left(b_{p k 1}, \dots, b_{p k K_p} \right)^\top.
\end{equation}
We have, for $p = 1, \dots, P$,
\begin{align*}
    \left(\Gamma \phi_k\right)^{(p)}(\cdot) &= \sum_{q = 1}^P \int_{\TT{q}} C_{p, q}(\cdot, s_q)\phi_k^{(q)}(s_q) \dd s_q \\
    &= \sum_{q = 1}^P \int_{\TT{q}} \Psi(\cdot)^{(p) \top} \mathbf{C}^{(p) \top} \left(\text{diag}\{\pi_1, \dots, \pi_N\} - \Pi\Pi^\top\right)\mathbf{C}^{(q)} \Psi^{(q)}(s_q) \Psi^{(q)}(s_q)^\top b_{q k} \dd s_q \\
    &= \Psi(\cdot)^{(p) \top} \mathbf{C}^{(p) \top} \left(\text{diag}\{\pi_1, \dots, \pi_N\} - \Pi\Pi^\top\right)\sum_{q = 1}^P \mathbf{C}^{(q)} \int_{\TT{q}} \Psi^{(q)}(s_q) \Psi(s_q)^{(q) \top} \dd s_q b_{q k} \\
    &= \Psi(\cdot)^{(p) \top} \mathbf{C}^{(p) \top} \left(\text{diag}\{\pi_1, \dots, \pi_N\} - \Pi\Pi^\top\right) \sum_{q = 1}^P \mathbf{C}^{(q)} \mathbf{W}^{(q)} b_{q k}. \\
\end{align*}
This equation is true for all $p = 1, \cdots, P$, this can be rewritten with matrices as
\begin{equation}
    \Gamma \phi_k(\cdot) = \Psi(\cdot)^{\top} \mathbf{C}^{\top} \left(\text{diag}\{\pi_1, \dots, \pi_N\} - \Pi\Pi^\top\right) \mathbf{C} \mathbf{W} b_{k}.
\end{equation}
From the eigenequation, we have that
\begin{equation}
    \Gamma \phi_k(\cdot) = \lambda_k \phi_k(\cdot) \Longleftrightarrow \Psi(\cdot)^{\top} \mathbf{C}^{\top} \left(\text{diag}\{\pi_1, \dots, \pi_N\} - \Pi\Pi^\top\right) \mathbf{C} \mathbf{W} b_{k} = \lambda_k \Psi(\cdot)^\top b_k.
\end{equation}
Since this equation must be true for all $t_p \in \TT{p}$, this imply the equation
\begin{equation}\label{eq:eigen_decom}
    \mathbf{C}^{\top} \left(\text{diag}\{\pi_1, \dots, \pi_N\} - \Pi\Pi^\top\right) \mathbf{C} \mathbf{W} b_{k} = \lambda_k b_k.
\end{equation}
As the eigenfunctions are assumed to be normalized, $\normH{\phi_k}^2 = 1$. And so, $b_k^\top \mathbf{W} b_k = 1$. Let $u_k = \mathbf{W}^{1/2}b_k$. Then, from \eqref{eq:eigen_decom}, we obtain
\begin{equation}\label{eq:eigen_cov_op}
    \mathbf{W}^{1/2} \mathbf{C}^{\top} \left(\text{diag}\{\pi_1, \dots, \pi_N\} - \Pi\Pi^\top\right) \mathbf{C} \mathbf{W}^{1/2} u_k = \lambda_k u_k \Longleftrightarrow \mathbf{A}^\top\mathbf{A} u_k = \lambda_k u_k.
\end{equation}
From the eigendecomposition of the matrix $M$, we get
\begin{equation}\label{eq:eigen_inner_prod}
    \mathbf{M}\boldsymbol{u}_k = l_k \boldsymbol{u}_k \Longleftrightarrow \mathbf{A}\mathbf{A}^\top \boldsymbol{u}_k = l_k \boldsymbol{u}_k.
\end{equation}
The equations \eqref{eq:eigen_cov_op} and \eqref{eq:eigen_inner_prod} are eigenequations in the classical PCA case, with the duality $X^\top X$ and $XX^\top$. Following \cite{pagesMultipleFactorAnalysis2014,hardleAppliedMultivariateStatistical2019}, we find that, for $1 \leq k \leq K$,
\begin{equation}
    \lambda_k = l_k, \quad \boldsymbol{u}_k = \frac{1}{\sqrt{l_k}}\mathbf{A} u_k \quad\text{and}\quad u_k = \frac{1}{\sqrt{l_k}} \mathbf{A}^\top \boldsymbol{u}_k.
\end{equation}
And finally, to get the coefficient of the eigenfunctions, for $1 \leq k \leq K$,
\begin{equation}
    b_k = \mathbf{W}^{-1/2}u_k = \frac{1}{\sqrt{l_k}} \mathbf{C}^\top \left(\mathrm{I}_{\!N} - \Pi\mathbf{1}_{\!N}^\top\right) \text{diag}\{\sqrt{\pi_1}, \dots, \sqrt{\pi_N}\}\boldsymbol{u}_k.
\end{equation}



\bibliographystyle{apalike}
\bibliography{biblio}

\makeatletter\@input{ms_ref.tex}\makeatother